\documentclass[]{alibaba-nlp}

\usepackage{url}
\usepackage{amsmath}
\usepackage{amssymb}
\usepackage{dsfont}
\usepackage{booktabs}
\usepackage{multirow}
\usepackage{graphicx}

\newcommand{\benchmark}{\textsc{VisDocAgentBench}}

\title{VisDocAgentBench: Benchmarking Agents for Visually Rich Document Retrieval}
\author[1,\dagger]{Lexiang Hu}
\author[]{Yanzhao Zhang}
\author[]{Mingxin Li}
\author[]{Dingkun Long}
\author[1]{Yikang Li}
\author[]{Fuwei Zhang}
\author[1]{Yisen Wang}
\author[1]{Zhouchen Lin}

\affiliation[1]{State Key Lab of General AI, School of Intelligence Science and Technology, Peking University}

\abstract{
Visually rich documents encode relevance through language, layout, structured visual elements, and corpus context, yet retrieval is typically evaluated by one-shot query--page matching. Agentic-search benchmarks usually score downstream question answering or report generation, leaving document ranking under iterative evidence acquisition underexplored. We introduce \benchmark, a closed-corpus benchmark comparing static and agentic retrieval under a shared ranked-output contract. It contains 2,375 pages from 100 documents and 120 unique-target queries balanced across direct, one-bridge, and two-bridge evidence structures. Relation-preserving construction yields semantic, relational, and visual queries, followed by full-document review and hard-negative validation. A strong late-interaction visual retriever reaches 97.50\% Recall@1 on direct items but 2.50\% on two-bridge items, exposing the limits of query--target matching when relevance depends on corpus context. Agents recover much of this loss, but planner choice and retrieval representation remain decisive. Every planner performs better with visual retrieval, whose best R@1 reaches 67.50\% versus 37.50\% for OCR-text. Ablations identify iterative search and page inspection as consequential capabilities, and providing the complete support context improves ranking on both routes. Trace analysis localizes the remaining losses to target discovery, candidate examination, and evidence-role integration. These findings motivate retrieval agents that combine modality-preserving discovery with evidence-directed verification.
}

\date{\today}
\contact{Lexiang Hu (\email{hulx@stu.pku.edu.cn})}
\correspondingauthor{Zhouchen Lin (\email{zlin@pku.edu.cn})}
\projectpage{\url{https://hulx2002.github.io/VisDocAgentBench}}

\begin{document}
\maketitle
\begingroup
\renewcommand{\thefootnote}{\fnsymbol{footnote}}
\footnotetext[2]{Work done during internship at Token Foundry, Alibaba Group.}
\endgroup

\section{Introduction}
\label{sec:introduction}

Visual retrieval has expanded from matching text to photographs and other natural images toward a common interface for heterogeneous multimodal content. Large-scale vision--language pretraining and dual-encoder objectives made open-vocabulary image matching scalable \citep{radford2021clip,jia2021align,zhai2022lit,zhai2023siglip}; instruction-aware and unified representations subsequently broadened retrieval across tasks, intents, and media \citep{wei2024uniir,zhang2024magiclens,jiang2025vlm2vec,lin2025mmembed,meng2025vlm2vecv2}. Visually rich documents fall within this broader interface, but they exhibit a different relevance structure. Whereas natural-image relevance is often expressed through depicted entities, attributes, actions, and scenes, documents are designed information artifacts in which language, two-dimensional layout, tables, diagrams, formulas, and interface elements jointly encode meaning \citep{faysse2025colpali,dong2025mmdocir,osmulski2025miraclvision,shorten2026irpapers,yan2026vdrsurvey}. Identifying a target unit may therefore require both reading its local text--layout composition and connecting it to definitions, results, or conditions elsewhere in the collection \citep{ma2024mmlongbenchdoc,chen2024documenthaystacks,cho2024m3docrag,wang2025vidorag}.

Retrieval is also shifting from one-shot ranking toward agentic search. Agents can move beyond fixed-request candidate scores by reformulating searches, inspecting results, invoking specialized tools, and accumulating evidence across steps \citep{yao2023react,nakano2022webgpt,jin2025searchr1,jiang2025mmsearch}. Work on visually rich documents, however, remains largely understanding-oriented, with retrieval used mainly for fixed-query page ranking \citep{faysse2025colpali,dong2025mmdocir,osmulski2025miraclvision,shorten2026irpapers} or to support document question answering \citep{mathew2021docvqa,ma2024mmlongbenchdoc,ouyang2025omnidocbench,wang2025vidorag,wang2026agenticocr}. Many recent agentic visual-search benchmarks support iterative interaction but focus primarily on natural images \citep{deng2026deepimagesearch,he2026vistahop,zhang2026visualseeker}. The missing intersection is agentic search over a closed visual-document corpus with direct evaluation of the final document ranking.

\begin{figure}[htbp]
  \centering
  \includegraphics[width=\linewidth]{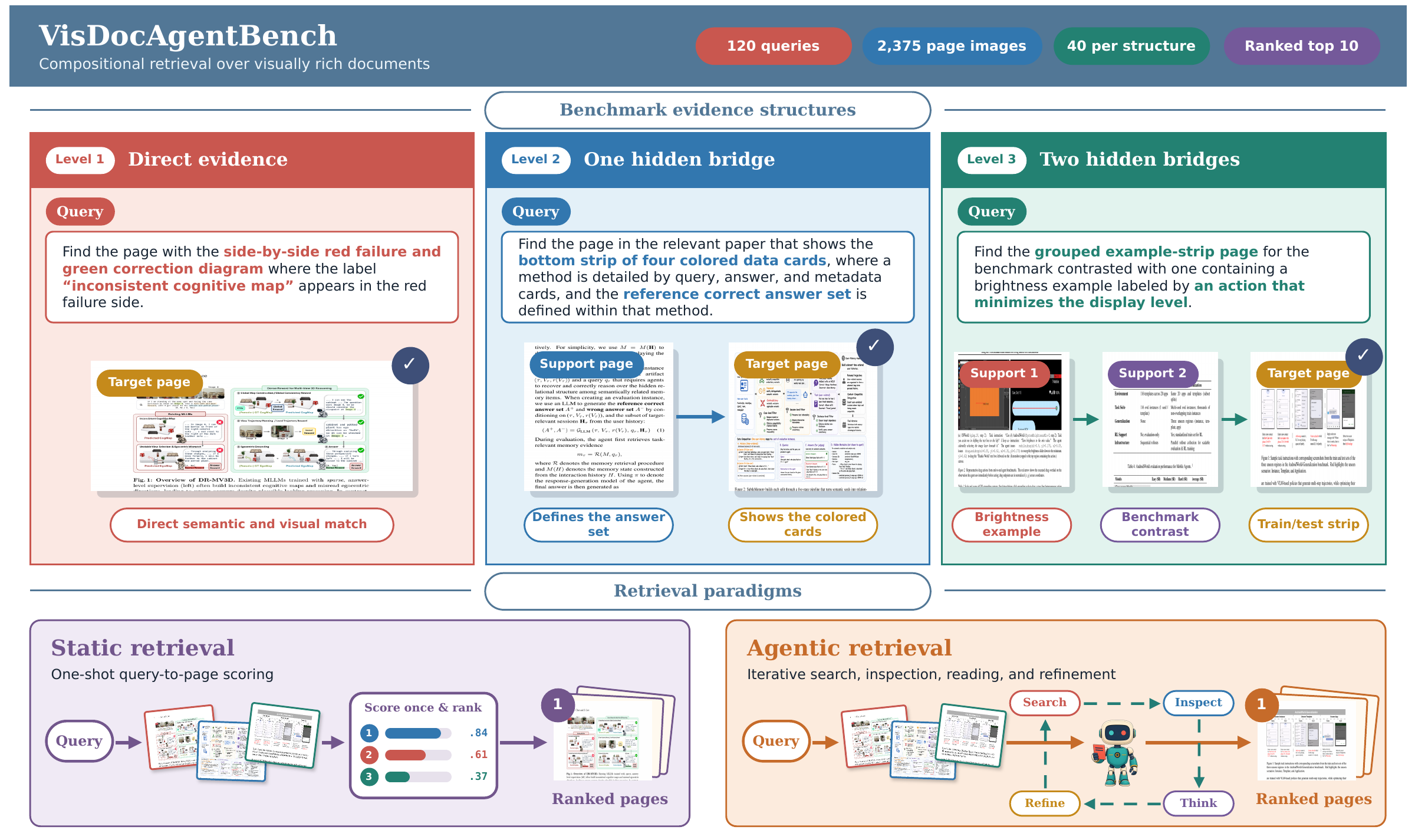}
  \caption{Overview of \benchmark{}. The upper panel depicts direct, one-bridge, and two-bridge evidence structures. The lower panel contrasts one-shot static scoring with iterative search and inspection. Construction annotations and gold labels are hidden from evaluated systems.}
  \label{fig:benchmark-task}
\end{figure}

We introduce \benchmark{}, a closed-corpus benchmark for ranked visual-document retrieval. Each query combines semantic, relational, and visual conditions and has one target page in the image corpus. Static retrievers and tool-using agents share the same ranked-output contract. The 120 items are balanced across direct, one-bridge, and two-bridge evidence structures, as illustrated in Figure~\ref{fig:benchmark-task}.

The benchmark contains 2,375 pages from 100 scientific documents. Its queries undergo relation-preserving construction, full-document review, and hard-negative validation. A late-interaction visual retriever is nearly perfect on direct items but falls from 97.50\% to 2.50\% Recall@1 on two-bridge items, showing that stronger query--page matching does not resolve distributed evidence. Tool-using agents recover much of this loss, although their wide variation under the same interface shows that tool access alone is insufficient. Every planner performs better with visual retrieval. Ablations identify iterative search and page inspection as consequential capabilities, while complete support context improves L2 and L3 ranking on both routes. The remaining failures concentrate in target discovery, candidate examination, and evidence-role integration. Together, these results call for visual-document representations and evidence-directed search policies to be developed jointly.

Our main contributions are as follows:
\begin{itemize}
  \item \textbf{Task formulation.} We formulate closed-corpus agentic visual-document retrieval with ranked retrieval as the endpoint and a shared output contract across retrieval paradigms.
  \item \textbf{Benchmark.} We release \benchmark{}, spanning a 2,375-page document corpus and 120 balanced, unique-target queries grounded in reviewed full-document evidence.
  \item \textbf{Construction.} We introduce a role-specific, relation-preserving construction pipeline with automatic screening, full-document review, constrained authoring, and hard-negative validation.
  \item \textbf{Evaluation.} We benchmark static retrievers and tool-using agents across visual and OCR-text routes, using controlled diagnostics and auditable traces to expose representation, search, and evidence-integration bottlenecks.
\end{itemize}

\section{Related Work}
\label{sec:related}

\subsection{Visual Retrieval over Images and Documents}

Large-scale vision--language pretraining supports efficient open-vocabulary image retrieval through dual encoders and related contrastive objectives \citep{radford2021clip,zhai2023siglip}. Instruction-aware and general-purpose retrievers extend this interface across tasks and media \citep{wei2024uniir,jiang2025vlm2vec,lin2025mmembed,zhou2025megapairs,meng2025vlm2vecv2,li2026qwen3vlembedding,hu2026agffembed}. Although these representations broaden query expressivity, their standard protocol encodes a request once and ranks candidates independently.

Visual-document retrieval must additionally preserve language, layout, tables, figures, and fine-grained regions. Existing approaches use direct page embeddings, late interaction, query-conditioned representations, region supervision, and staged visual refinement \citep{faysse2025colpali,yu2025visrag,cho2024m3docrag,yang2026realign,abdallah2026argusretriever,guan2026lightstar}. Complementary work improves contextualization and index efficiency \citep{yan2026visuallatechunking,yan2026prunethenmerge,qin2026multivectorcompression,georgiou2026hydra}, while document benchmarks compare page-, layout-, multilingual-, and large-corpus retrieval settings \citep{dong2025mmdocir,osmulski2025miraclvision,chen2024documenthaystacks,shorten2026irpapers,yan2026vdrsurvey}. Together, these methods and benchmarks establish the representational challenges of visual-document retrieval, primarily under fixed-query ranking protocols. \benchmark{} extends this line by evaluating systems that can iteratively search and inspect the corpus before producing the final visual-unit ranking.

\subsection{Agentic Search}

Language-model agents interleave reasoning and external actions \citep{yao2023react}. Search agents apply this pattern to query reformulation, iterative retrieval, and evidence accumulation across browsers and search engines \citep{nakano2022webgpt,jin2025searchr1,gou2025mind2web2,wei2025browsecomp,du2025deepresearchbench}. Multimodal variants add visual queries and observations \citep{jiang2025mmsearch,li2025mmbrowsecomp}, while recent learned agents target longer-horizon and actively visual search \citep{zhang2026vsearcher,chen2026opensearchvl,peng2026mtaagent,liu2026pointsseeker,zhang2026visualseeker}. Their retrieval actions usually support a generated answer, report, or task outcome, leaving the ranking trajectory itself only indirectly measured.

Process-oriented benchmarks increasingly expose intermediate visual-search behavior through checkpoints, cropping, visual browsing, and multimodal evidence traces \citep{wei2026agenticmme,he2026vistahop,zhang2026visbrowsebench,huang2026mmdeepresearchbench}. Document-focused agents similarly combine retrieval with selective reading or OCR for question answering and document understanding \citep{wang2025vidorag,wang2026agenticocr,wang2026docarena}. DeepImageSearch instead evaluates context-aware image retrieval as the endpoint over personal visual histories \citep{deng2026deepimagesearch}. \benchmark{} shares the emphasis on auditable interaction but targets ranked retrieval from a common closed corpus of visually rich documents.

\subsection{Benchmarks for Visual and Agentic Retrieval}

Visual-document benchmarks span image-based question answering and chart reasoning \citep{mathew2021docvqa,masry2022chartqa}, long-context and multi-document understanding \citep{ma2024mmlongbenchdoc,chen2024documenthaystacks}, document parsing \citep{ouyang2025omnidocbench}, and direct page retrieval \citep{dong2025mmdocir,osmulski2025miraclvision,shorten2026irpapers}. Evaluation in agentic settings instead spans browser actions, difficult information discovery, multimodal answers, task completion, and retrieval for agentic coding \citep{koh2024visualwebarena,jiang2025mmsearch,li2025mmbrowsecomp,wei2025browsecomp,zhang2026corebench}. \benchmark{} connects these lines through a shared ranked-page endpoint and exposes the search, inspection, modality-conversion, and final-ranking trajectory.

Recent multimodal benchmarks also emphasize shortcut control, structured quality assurance, and post-construction verification \citep{yue2025mmmupro,chen2025megabench,wei2025browsecomp,deng2026deepimagesearch}. We adapt these principles to compositional retrieval by validating relation paths, reviewing surviving items in full-document context, constraining which clues enter the query, and auditing hard negatives returned by a strong visual retriever. The resulting benchmark separates corpus-grounded evidence structure from the retrieval paradigm used to solve it.

\section{Task Formulation}
\label{sec:task}

\subsection{Closed-Corpus Visual-Document Retrieval}

Visually rich documents encode linguistic content, spatial organization, and graphical structure on a shared visual surface. We call a retrievable item a \emph{visual document unit} $x_i$, which may combine embedded text, layout, tables, diagrams, formulas, screenshots, or interface components.

Let the fixed corpus be $\mathcal{C}=\{x_1,\ldots,x_N\}$. Given a natural-language query $q$ and a unique target $a_q\in\mathcal{C}$, a system returns an ordered list of distinct visual units
\begin{equation}
  R_q^K=[x_{(1)},\ldots,x_{(K)}].
  \label{eq:ranked-output}
\end{equation}
The objective is to rank $a_q$ as highly as possible. Relevance may jointly depend on language, concept relations, and visual-layout properties, while evaluation concerns the target ranking.

\subsection{Context-Dependent Compositional Relevance}

We distinguish \emph{target-local evidence}, visible on a candidate, from \emph{corpus-context evidence}, located in other units and useful for resolving semantic or relational conditions in $q$. A locally plausible unit can remain incorrect when it fails the complete corpus-grounded relation. Context elsewhere in the corpus can disambiguate such candidates.

For analysis, let $E_q\subseteq\mathcal{C}\setminus\{a_q\}$ denote a possibly empty latent support set such that
\begin{equation}
  \operatorname{Rel}(q,a_q\mid E_q)=1,
\end{equation}
where $\operatorname{Rel}$ indicates whether the visual unit satisfies the query under the contextual evidence. Target-local evidence can establish relevance when $E_q=\varnothing$. Otherwise, distributed context can contribute to the judgment. Appendix~\ref{app:task-instantiation} specifies how $a_q$ and $E_q$ are instantiated in the benchmark and how target validity is adjudicated.

\subsection{Static and Agentic Retrieval}

A static retriever independently scores each query--unit pair with a function $s(q,x_i)$ and produces one fixed ordering:
\begin{equation}
  R_{\mathrm{static}}(q)=\operatorname{argsort}_{i}\,s(q,x_i).
  \label{eq:static-ranking}
\end{equation}
Evaluation uses the first $K$ positions of this ordering.

Agentic retrieval is a budgeted action--observation process. Let $\mathcal{U}_{\mathrm{tool}}$ denote the available evidence-gathering actions, $\mathcal{I}$ the corpus interface, and $h_0=(q)$ the initial interaction history. For some $T\leq B$, the policy $\pi$ selects an evidence-gathering action at each step $t\in\{1,\ldots,T\}$, receives an observation $o_t$, and appends both to its history:
\begin{equation}
  \begin{cases}
    u_t=\pi(h_{t-1})\in\mathcal{U}_{\mathrm{tool}},\\
    o_t=\mathcal{I}(u_t,\mathcal{C}),\\
    h_t=h_{t-1}\oplus(u_t,o_t),
  \end{cases}
  \label{eq:agent-history}
\end{equation}
where $\oplus$ denotes sequence concatenation. After the interaction ends, an output policy maps the accumulated history to a ranking:
\begin{equation}
  R_{\mathrm{agent}}^K(q)=\pi_{\mathrm{out}}(h_T),\qquad T\leq B.
  \label{eq:agent-submission}
\end{equation}
Embedding search can serve as an agent tool. Agentic retrieval gains the additional capacity to condition later searches and the final ordering on corpus evidence observed during the episode.

\section{Benchmark Construction and Quality Control}
\label{sec:construction}

\begin{figure}[htbp]
  \centering
  \includegraphics[width=\linewidth]{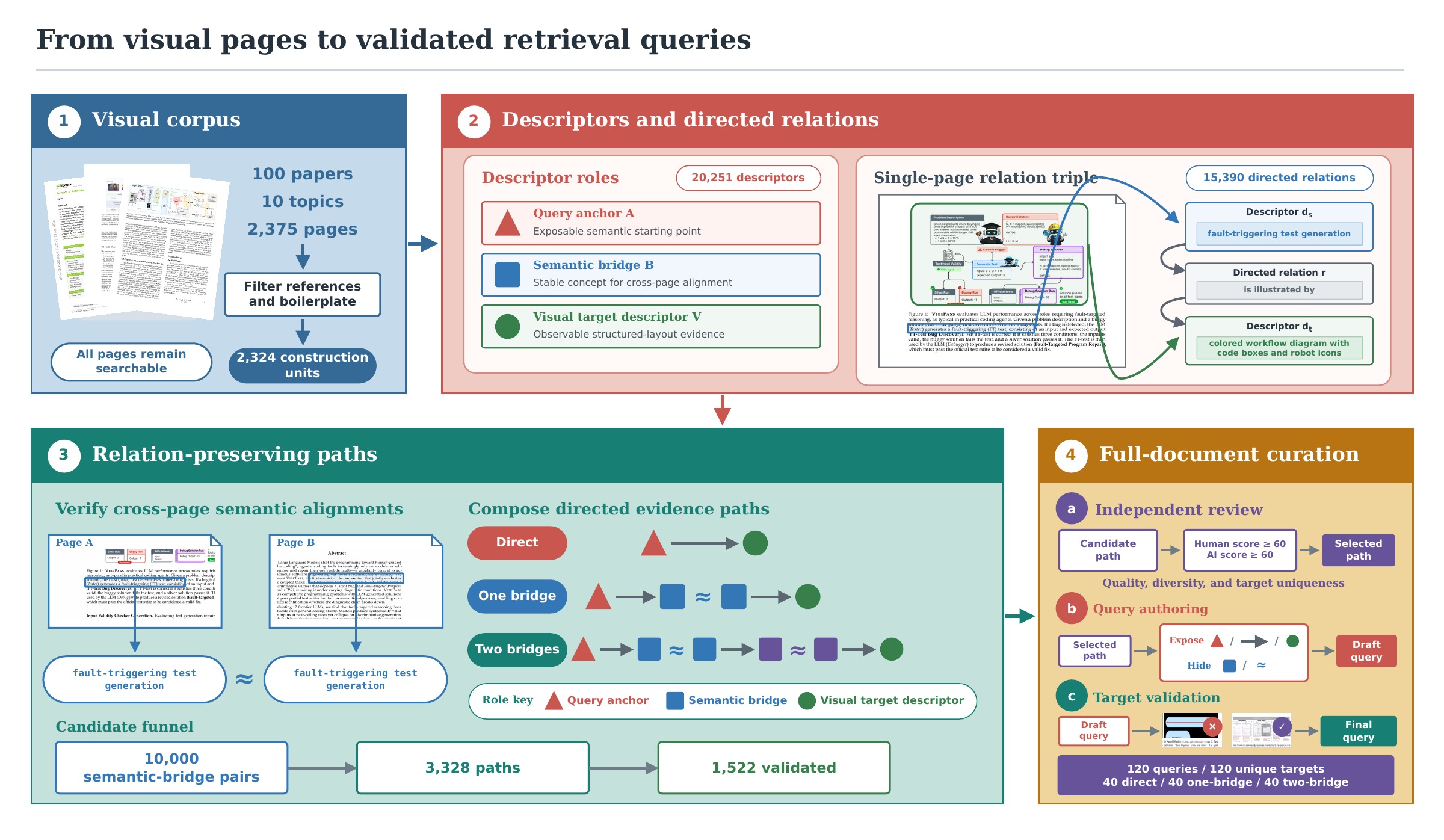}
  \caption{Construction and quality-control pipeline. Role-specific descriptors, directed relations, semantic alignments, and evidence paths supervise benchmark construction but are hidden during evaluation. The final benchmark is formed only after independent full-document scoring, constrained query authoring, and hard-negative target validation.}
  \label{fig:construction}
\end{figure}

Figure~\ref{fig:construction} summarizes the pipeline from visual-unit preparation through relation-preserving path construction and full-document quality control.

\subsection{Corpus Collection and Visual-Unit Preparation}

We use scientific papers because their pages densely combine text, layout, tables, and diagrams while distributing recurring concepts across units. They preserve traceable provenance and support deterministic rendering plus full-context verification, as established by prior multimodal retrieval and document-reasoning benchmarks \citep{chen2024documenthaystacks,cho2024m3docrag,dong2025mmdocir,shorten2026irpapers}.

We sample 100 arXiv papers published in 2026, taking 10 papers from each of 10 relatively distinct topics such as agent systems and world models. In \benchmark{}, each visual document unit is one rendered full page, giving 2,375 images with document and page provenance. We use 2,324 non-reference, content-bearing pages for query construction. All 2,375 images form the retrieval corpus.
Appendix~\ref{app:topic-selection} details topic and document selection and lists the complete source inventory in Table~\ref{tab:document-inventory}. Appendix~\ref{app:artifact-licensing} describes source-license verification, redistribution rules, and privacy safeguards.

\subsection{Role-Specific Descriptors and Directed Relations}

For each construction unit, a VLM extracts role-specific descriptors $d$ and directed relations $r$. Each descriptor is assigned one of three roles. A \emph{query anchor} provides a semantic starting point that can be expressed without identifying the target. A \emph{semantic bridge} denotes a stable concept that can align across units but remains hidden from the final query. A \emph{visual target descriptor} captures observable layout or structured evidence without relying on captions, identifiers, or answer-revealing text. A directed relation $r$ connects two compatible descriptors $d_s$ and $d_t$ in the same unit, forming a local triple $d_s\xrightarrow{r}d_t$. These local triples form the building blocks of relation-preserving paths.

Role and quality validation removes revealing query anchors, visual target descriptors that paraphrase ordinary body text, incompatible role pairs, and relations grounded in rejected descriptors. The extraction pass yields 21,955 descriptors and 17,693 relations. Validation retains 20,251 descriptors---6,149 query anchors, 10,531 semantic bridges, and 3,571 visual target descriptors---with 15,390 usable relations. Figure~\ref{fig:construction} summarizes this inventory, and Appendix~\ref{app:automatic-construction} details automated descriptor extraction and validation.

\subsection{Relation-Preserving Path Construction}

We embed validated semantic bridges with Qwen3-Embedding-8B \citep{zhang2025qwen3embedding} and retrieve candidate pairs independently within each topic, limiting accidental cross-domain similarity. Embedding similarity provides a high-recall semantic screen, after which a VLM verifies conceptual equivalence or contextual compatibility. The VLM accepts 4,708 of the 10,000 candidate pairs as \emph{semantic alignments} between bridges from different units. Within a document, stable mid-level concepts may align; across documents, we require proper names, aliases, or variants of the same named object.

Let $A$ denote a query anchor, $B_i$ a semantic bridge, $V$ a visual target descriptor, $r_i$ a directed intra-unit relation, and $\approx$ a verified cross-unit semantic alignment. We construct path templates:
\begin{equation}
  \label{eq:path-templates}
  \begin{cases}
    \mathrm{L1:} & A\xrightarrow{r_1}V, \\
    \mathrm{L2:} & A\xrightarrow{r_1}B_1\approx B_2\xrightarrow{r_2}V, \\
    \mathrm{L3:} & A\xrightarrow{r_1}B_1\approx B_2\xrightarrow{r_2}B_3
    \approx B_4\xrightarrow{r_3}V.
  \end{cases}
\end{equation}
L1 contains a direct relation from the query anchor to the visual target descriptor; L2 and L3 introduce one and two latent semantic alignments. Appendix~\ref{app:path-query-constraints} gives the complete level-specific path and query constraints.

The verified alignments compose 3,328 candidate paths. Automatic path validation checks semantic coherence, evidence necessity for L2 and L3, and visual necessity. It rejects paths that expose later descriptors, fail to form an interpretable relation chain, or lack grounding in a structured visual region. This leaves 1,522 paths: 364 L1, 1,058 L2, and 100 L3.
Appendix~\ref{app:automatic-construction} details the semantic-alignment screen, path-composition limits, and automatic-validation criteria.

\subsection{Full-Document Path Review and Selection}

Every automatically validated path is reviewed in the context of all involved source documents. A human reviewer and an AI reviewer independently score semantic validity, evidence necessity, bridge specificity, visual grounding, query rewritability, and diversity on a 0--100 scale. Semantic validity tests whether every local relation and cross-page alignment is supported by the papers. Evidence necessity tests whether the intermediate concepts are needed to resolve the endpoint. Bridge specificity excludes broad topical overlap, while visual grounding requires an observable structured region on the target. A path remains eligible only when both scores are at least 60 and is then ranked by their mean. We select 40 paths per level, prioritizing quality and using diversity to separate close candidates. Target deduplication and replacement produce 120 distinct targets.
Appendix~\ref{app:path-review} and Appendix~\ref{app:path-selection} detail full-document scoring, selection, and target deduplication.

\subsection{Final Query Authoring and Target Validation}

Final queries express the query anchor $A$, every directed relation $r_i$ in the path skeleton, and an abstracted visual target descriptor $V$, while hiding semantic bridges $B_i$ and their alignments $\approx$ behind natural generic references. Visual specificity is calibrated so that the visual phrase can match multiple units but its conjunction with semantic and relational constraints identifies the target. Queries request a visual unit without exposing source identifiers, exact captions, or construction terminology.
Table~\ref{tab:exposure-policy} in Appendix~\ref{app:path-query-constraints} summarizes the level-specific exposure and hiding policy.

Each final query is checked against every page of its involved source documents for plausible alternatives satisfying the complete request. Qwen3-VL-Embedding-8B \citep{li2026qwen3vlembedding} then retrieves ten candidates from the full corpus, and every non-gold result is inspected. A valid alternative triggers minimal disambiguation; an item is replaced when uniqueness cannot be restored naturally. The resulting query is rechecked against the relevant full documents and retrieved alternatives.
Appendix~\ref{app:query-authoring} and Appendix~\ref{app:hard-negative-validation} provide the full authoring and validation protocols.

\subsection{Benchmark Statistics}

\begin{table}[htbp]
  \centering
  \small
  \setlength{\tabcolsep}{4pt}
  \caption{Corpus, construction, and final benchmark statistics.}
  \label{tab:benchmark-stats}
  \begin{tabular}{lr}
    \toprule
    \textbf{Statistic} & \textbf{Count} \\
    \midrule
    \multicolumn{2}{l}{\textit{Corpus}} \\
    Source documents / topics & 100 / 10 \\
    Rendered visual units & 2,375 \\
    Query-construction units & 2,324 \\
    \midrule
    \multicolumn{2}{l}{\textit{Construction}} \\
    Candidate semantic-bridge pairs & 10,000 \\
    Constructed paths & 3,328 \\
    Automatically validated paths & 1,522 \\
    \midrule
    \multicolumn{2}{l}{\textit{Final benchmark}} \\
    Queries / unique targets & 120 / 120 \\
    L1 / L2 / L3 queries & 40 / 40 / 40 \\
    Single-unit / same-doc. / cross-doc. & 40 / 74 / 6 \\
    Documents represented in selected paths & 55 \\
    \bottomrule
  \end{tabular}
\end{table}

Table~\ref{tab:benchmark-stats} summarizes the construction funnel and final composition. Final queries average 25.6 words (median 26; range 15--39). Seventy-four evidence paths remain within one document and six cross document boundaries. Table~\ref{tab:topic-funnel} in Appendix~\ref{app:topic-funnel-details} reports the topic-wise funnel.

\section{Experimental Setup}
\label{sec:setup}

\subsection{Evaluation Protocol and Metrics}

The evaluation corpus contains $N=2{,}375$ units. Each query has one gold target, and systems are evaluated under a contract requiring $K=10$ distinct opaque identifiers. Table~\ref{tab:annotation-fields} in Appendix~\ref{app:annotation-fields} specifies annotation roles and their inference-time visibility. We report Recall@$k$ for $k\in\{1,3,5,10\}$ and MRR@10. Let $r_q=\operatorname{rank}_q(a_q)$, with missing or invalid rankings assigned $r_q=\infty$:
\begin{equation}
 \begin{cases}
  \displaystyle \mathrm{R@}k = \frac{1}{|Q|}\sum_{q\in Q}\mathds{1}[r_q\leq k],\\[4pt]
  \displaystyle \mathrm{MRR@10} = \frac{1}{|Q|}\sum_{q\in Q}\frac{\mathds{1}[r_q\leq 10]}{r_q}.
 \end{cases}
\end{equation}
R@1 is the primary measure, R@10 measures final top-10 coverage, and MRR@10 summarizes target ordering within that list. We report overall and level-wise results.

\subsection{Baselines and Implementation Details}
\label{sec:baselines-implementation}

\paragraph{Static retrieval.}
Our single-vector baselines encode page images with Qwen3-VL-Embedding-8B \citep{li2026qwen3vlembedding} or page OCR with Qwen3-Embedding-8B \citep{zhang2025qwen3embedding}, ranking all pages by cosine similarity. OCR is produced once with PaddleOCR-VL-1.6 and PP-DocLayoutV3 \citep{zhang2026paddleocrvl16}. We add BM25 \citep{robertson2009bm25} over the same OCR records and reciprocal-rank fusion of BM25 with the dense OCR ranking \citep{cormack2009rrf}. Nemotron-ColEmbed-VL-8B-V2 \citep{moreira2026nemotron} provides a stronger visual-document baseline through token-level late interaction. Retrieval parameters follow the published defaults and are not tuned on \benchmark{}. Appendix~\ref{app:embedding-indices} gives the index and scoring details.

\paragraph{Tool-using retrieval.}
Both agent routes reuse the matching single-vector index but allow the planner to issue multiple searches. The visual route supports visual search and on-demand OCR; the OCR-text route supports text search and cached page OCR. Both expose full-page inspection, batched regional cropping, and ranked submission, as summarized in Table~\ref{tab:main-tools}.

\begin{table}[htbp]
  \centering
  \small
  \setlength{\tabcolsep}{5pt}
  \caption{Information interface for the two tool-using retrieval routes. Table~\ref{tab:tool-schemas} in Appendix~\ref{app:agent-tools} gives the complete schemas and argument constraints.}
  \label{tab:main-tools}
  \begin{tabular}{lp{0.25\textwidth}p{0.57\textwidth}}
    \toprule
    \textbf{Route} & \textbf{Operation} & \textbf{Returned information} \\
    \midrule
    \multirow{2}{*}{Visual} & Visual search & Ranked opaque page handles and similarity scores. \\
      & On-demand OCR & Text and layout blocks from one discovered page or crop. \\
    \midrule
    \multirow{2}{*}{OCR-Text} & Text search & Ranked opaque page handles and similarity scores; no snippets. \\
      & Cached page OCR & Full OCR text for one discovered page. \\
    \midrule
    \multirow{3}{*}{Both} & Page inspection & Full-page images for up to ten discovered pages. \\
      & Regional cropping & Up to ten crops from previously inspected pages. \\
      & Ranked submission & Up to ten distinct discovered pages; exactly ten when available. \\
    \bottomrule
  \end{tabular}
\end{table}

\paragraph{Planner models and interaction budget.}
We evaluate GPT-5.5, GPT-5.6-luna, GPT-5.6-terra, GPT-5.6-sol, Claude Opus 4.8, Claude Fable 5, Claude Sonnet 5, Claude Opus 5, and the open-weight Qwen3.5-397B-A17B with thinking enabled and disabled. Every agent receives up to 12 interaction steps, followed when needed by a ranking-only finalization call, and is evaluated under the same top-10 contract as the non-agentic baselines. For GPT-5.6-sol, we ablate iterative search, page inspection, route-specific text access, and regional cropping. Appendices~\ref{app:planner-configuration} and~\ref{app:ablation-configurations} specify the planner protocol and ablation configurations. Table~\ref{tab:gpt56-resource} in Appendix~\ref{app:runtime-accounting} reports per-query resource indicators. Appendix~\ref{app:open-weight-resources} describes the open-weight deployment and reports per-setting runtimes in Table~\ref{tab:qwen-runtime}.

\paragraph{Controlled evidence intervention.}
For L2 and L3, we rerun GPT-5.6-sol after presenting every annotated support
page as an unlabeled initial page observation.  The query, corpus, tools, and
12-step interaction budget remain fixed, yielding a paired comparison with the standard
agent run in Section~\ref{sec:behavior}.

\section{Results and Analysis}
\label{sec:results}

\subsection{Main Results}

Table~\ref{tab:main-results} compares static retrievers and tool-using agents
under the same ranked-output contract.  The visual and OCR-text routes operate
over the same corpus.  Complete level-wise results appear in
Table~\ref{tab:level-results} of Appendix~\ref{app:level-main-results};
Table~\ref{tab:gpt56-topic-results} in Appendix~\ref{app:topic-results} reports
the GPT-5.6-sol topic breakdown.

\begin{table}[htbp]
  \centering
  \small
  \setlength{\tabcolsep}{3.0pt}
  \caption{Overall retrieval results on \benchmark{} (\%). Metrics use all 120 queries, with invalid completed episodes scored as zero. Bold marks the best value in each route and metric.}
  \label{tab:main-results}
  \begin{tabular}{lrrrrrrrrrr}
    \toprule
    \multirow{2}{*}{\textbf{Retriever / Planner}} &
    \multicolumn{5}{c}{\textbf{Visual}} & \multicolumn{5}{c}{\textbf{OCR-Text}} \\
    \cmidrule(lr){2-6}\cmidrule(lr){7-11}
      & \textbf{R@1} & \textbf{R@3} & \textbf{R@5} & \textbf{R@10} & \textbf{MRR@10}
      & \textbf{R@1} & \textbf{R@3} & \textbf{R@5} & \textbf{R@10} & \textbf{MRR@10} \\
    \midrule
    \multicolumn{11}{l}{\textbf{Static Retrievers}} \\
    \midrule
    Qwen3 Embedding
      & 20.83 & 30.00 & 34.17 & 45.00 & 27.34
      & 1.67 & 5.83 & 13.33 & 19.17 & 6.01 \\
    BM25
      & -- & -- & -- & -- & --
      & 6.67 & 15.00 & 19.17 & 30.00 & 12.46 \\
    BM25 + dense RRF
      & -- & -- & -- & -- & --
      & 5.00 & 14.17 & 19.17 & 26.67 & 11.41 \\
    Nemotron ColEmbed
      & 40.00 & 52.50 & 65.00 & 70.00 & 48.86
      & -- & -- & -- & -- & -- \\
    \midrule
    \multicolumn{11}{l}{\textbf{Tool-Using Agents: Closed-Source Planners}} \\
    \midrule
    GPT-5.5
      & 60.00 & 61.67 & 64.17 & 69.17 & 61.87
      & 26.67 & 27.50 & 27.50 & 28.33 & 27.19 \\
    GPT-5.6-luna
      & 43.33 & 47.50 & 50.83 & 57.50 & 46.75
      & 12.50 & 14.17 & 18.33 & 21.67 & 14.77 \\
    GPT-5.6-terra
      & 54.17 & 57.50 & 58.33 & 64.17 & 56.65
      & 26.67 & 28.33 & 28.33 & 33.33 & 27.92 \\
    GPT-5.6-sol
      & 61.67 & 65.00 & 65.83 & 68.33 & 63.58
      & 36.67 & 38.33 & 40.83 & 44.17 & 38.40 \\
    Claude Opus 4.8
      & 47.50 & 54.17 & 59.17 & 68.33 & 52.39
      & 24.17 & 27.50 & 30.00 & 40.83 & 27.38 \\
    Claude Fable 5
      & 62.50 & 67.50 & \textbf{71.67} & \textbf{80.00} & 66.69
      & 35.00 & \textbf{41.67} & 43.33 & \textbf{52.50} & 39.34 \\
    Claude Sonnet 5
      & 30.00 & 40.83 & 46.67 & 55.83 & 37.33
      & 17.50 & 20.00 & 22.50 & 23.33 & 19.12 \\
    Claude Opus 5
      & \textbf{67.50} & \textbf{70.83} & \textbf{71.67} & 75.00 & \textbf{69.43}
      & \textbf{37.50} & \textbf{41.67} & \textbf{45.00} & 49.17 & \textbf{40.57} \\
    \midrule
    \multicolumn{11}{l}{\textbf{Tool-Using Agents: Open-Weight Planner}} \\
    \midrule
    Qwen3.5 (thinking)
      & 27.50 & 33.33 & 35.00 & 40.00 & 31.20
      & 8.33 & 9.17 & 10.00 & 15.83 & 9.70 \\
    Qwen3.5 (no thinking)
      & 19.17 & 23.33 & 30.00 & 38.33 & 23.55
      & 1.67 & 4.17 & 7.50 & 10.00 & 3.83 \\
    \bottomrule
  \end{tabular}
\end{table}

\paragraph{Stronger static matching exposes the evidence-path gap.}
Nemotron-ColEmbed-VL-8B-V2 is the strongest static retriever, reaching 40.00\% visual
R@1 overall.  Table~\ref{tab:level-results} in
Appendix~\ref{app:level-main-results} reveals a sharper level-wise contrast:
R@1 is 97.50\% on direct items and 2.50\% on two-bridge items.  A
late-interaction retriever can
therefore solve nearly every direct visual match while still failing when the
query--target relation depends on evidence elsewhere in the corpus.  OCR-text
retrieval remains weaker under dense, sparse, and fused rankings, with BM25
providing its strongest static R@1 at 6.67\%.

\paragraph{Interaction helps, but planner choice remains decisive.}
Agents using the same tools and indices vary widely: across the evaluated
planners, visual R@1 ranges from 19.17\% to 67.50\%, while OCR-text R@1 ranges
from 1.67\% to 37.50\%.  Claude Opus 5 achieves the strongest R@1 on both
routes, while Claude Fable 5 retrieves the most targets within the top 10.
The different leaders show that finding the target and placing it first require
distinct planner capabilities.  Iterative search and inspection can therefore
recover failures of static ranking, but only when the planner turns its
observations into better searches and final rankings.

\paragraph{Visual retrieval is consistently stronger than OCR-text retrieval.}
Every planner performs better on the visual route.  The R@1 difference ranges
from 12.50 points for Claude Sonnet 5 (30.00\% visual versus 17.50\% OCR-text)
to 33.33 points for GPT-5.5 (60.00\% versus 26.67\%).  This gap indicates that
full-page visual representations preserve layout and structured visual evidence
that an OCR-only index may discard during candidate discovery.  The gap arises
during global search, while targeted text extraction remains useful for
verifying selected pages.

\paragraph{Longer evidence paths remain challenging under interaction.}
As detailed in Table~\ref{tab:level-results} of
Appendix~\ref{app:level-main-results}, agentic interaction raises performance
but does not eliminate the path-length effect.  GPT-5.6-sol R@1 declines from
85.00\% to 40.00\% on the visual route and from 42.50\% to 27.50\% on the
OCR-text route, while Claude Opus 5 declines from 92.50\% to 47.50\% on the
visual route, indicating that longer evidence paths remain challenging even
under iterative search and inspection.
Table~\ref{tab:scope-results} in Appendix~\ref{app:path-scope-results} provides
the complementary GPT-5.6-sol breakdown by single-unit, same-document, and
cross-document paths.

\subsection{Ablation Studies}
\label{sec:ablations}

We ablate iterative search, page inspection, text access, and regional cropping
for GPT-5.6-sol while keeping the remaining settings fixed.  The
no-iterative-search setting applies VLM reranking to a fixed candidate set
containing the 30 highest-ranked pages from the route-specific retriever.  This
size slightly exceeds the full agent's observed mean page exposure on both
routes: valid episodes examine 25.22 pages with visual search and 28.22 with
OCR-text search (Table~\ref{tab:gpt56-resource} in
Appendix~\ref{app:runtime-accounting}).  It therefore provides a comparable
verification budget across routes.  No inspection also removes dependent
cropping.
Table~\ref{tab:ablations} gives the overall results, and
Table~\ref{tab:level-ablation} in Appendix~\ref{app:level-ablations} reports
every level.

\begin{table}[htbp]
  \centering
  \small
  \setlength{\tabcolsep}{2.0pt}
  \caption{GPT-5.6-sol ablations on all 120 queries (\%). $^\dagger$No-inspection also removes dependent regional cropping.}
  \label{tab:ablations}
  \begin{tabular}{lrrrrr}
    \toprule
    \textbf{Configuration} & \textbf{R@1} & \textbf{R@3} & \textbf{R@5} & \textbf{R@10} & \textbf{MRR@10} \\
    \midrule
    \multicolumn{6}{l}{\textbf{Visual}} \\
    \midrule
    Full agent & 61.67 & 65.00 & 65.83 & 68.33 & 63.58 \\
    No iterative search & 53.33 & 59.17 & 59.17 & 60.83 & 56.35 \\
    No regional crop & 64.17 & 69.17 & 71.67 & 74.17 & 67.34 \\
    No OCR & 60.00 & 62.50 & 66.67 & 69.17 & 62.40 \\
    No page inspection$^\dagger$ & 45.83 & 50.83 & 57.50 & 63.33 & 50.03 \\
    \midrule
    \multicolumn{6}{l}{\textbf{OCR-Text}} \\
    \midrule
    Full agent & 36.67 & 38.33 & 40.83 & 44.17 & 38.40 \\
    No iterative search & 27.50 & 29.17 & 30.83 & 30.83 & 28.61 \\
    No regional crop & 34.17 & 37.50 & 38.33 & 40.00 & 35.97 \\
    No page OCR & 33.33 & 35.00 & 35.83 & 39.17 & 34.63 \\
    No page inspection$^\dagger$ & 15.00 & 19.17 & 23.33 & 27.50 & 18.28 \\
    \bottomrule
  \end{tabular}
\end{table}

\paragraph{Iterative search improves both retrieval routes.}
Removing iterative search lowers R@1 by 8.34 points on the visual route and
9.17 points on OCR-text.  The ablation uses 11 VLM calls and 30 page
observations (Table~\ref{tab:gpt56-resource} in
Appendix~\ref{app:runtime-accounting}), yet its frozen candidate set
prevents observations from guiding subsequent retrieval.  Repeated candidate
inspection therefore does not substitute for search conditioned on accumulated
evidence.

\paragraph{Direct page inspection provides the largest gain.}
Among all ablations, removing page inspection causes the largest R@1 losses,
with drops of 15.84 and 21.67 points on the visual and OCR-text routes,
respectively.  These drops identify visual inspection of retrieved candidates
as the key verification step for determining whether semantic clues and visual
structure co-occur on the same page.

\paragraph{Text access exposes a path-dependent exploration--verification trade-off.}
Removing route-specific text access lowers overall R@1 by 1.67 points on the
visual route and 3.34 points on the OCR-text route.  The level-wise results in
Table~\ref{tab:level-ablation} of Appendix~\ref{app:level-ablations} show that
this aggregate benefit is concentrated on shorter paths: without text access,
R@1 drops by 7.50 and 10.00 points on visual L1 and L2, and by 2.50 and 10.00
points on OCR-text L1 and L2.  The direction reverses on L3, where R@1 rises by
12.50 and 2.50 points, respectively.  Trace-level tool usage shows that removing
text access shifts actions toward additional searches and full-page inspection.
Text therefore supports local verification, but current planners do not
reliably balance it against the broader corpus exploration required by longer
evidence paths.

\paragraph{Regional cropping has route-dependent effects.}
Removing cropping raises visual R@1 by 2.50 points and lowers OCR-text R@1 by
2.50 points, with mixed shifts across levels and metrics.  Visually rich pages
often require both local details, such as table cells or interface labels, and
their surrounding headers, rows, or diagrams.  A crop can clarify the former
while weakening the latter.  Current planners therefore do not reliably
determine when local detail outweighs the loss of page context.  These results
motivate the development of multimodal foundation models that can integrate
magnified evidence with its surrounding layout, thereby invoking cropping when
it is most informative.

\subsection{Evidence Intervention and Trace Diagnostics}
\label{sec:behavior}

\begin{figure}[htbp]
  \centering
  \includegraphics[width=\linewidth]{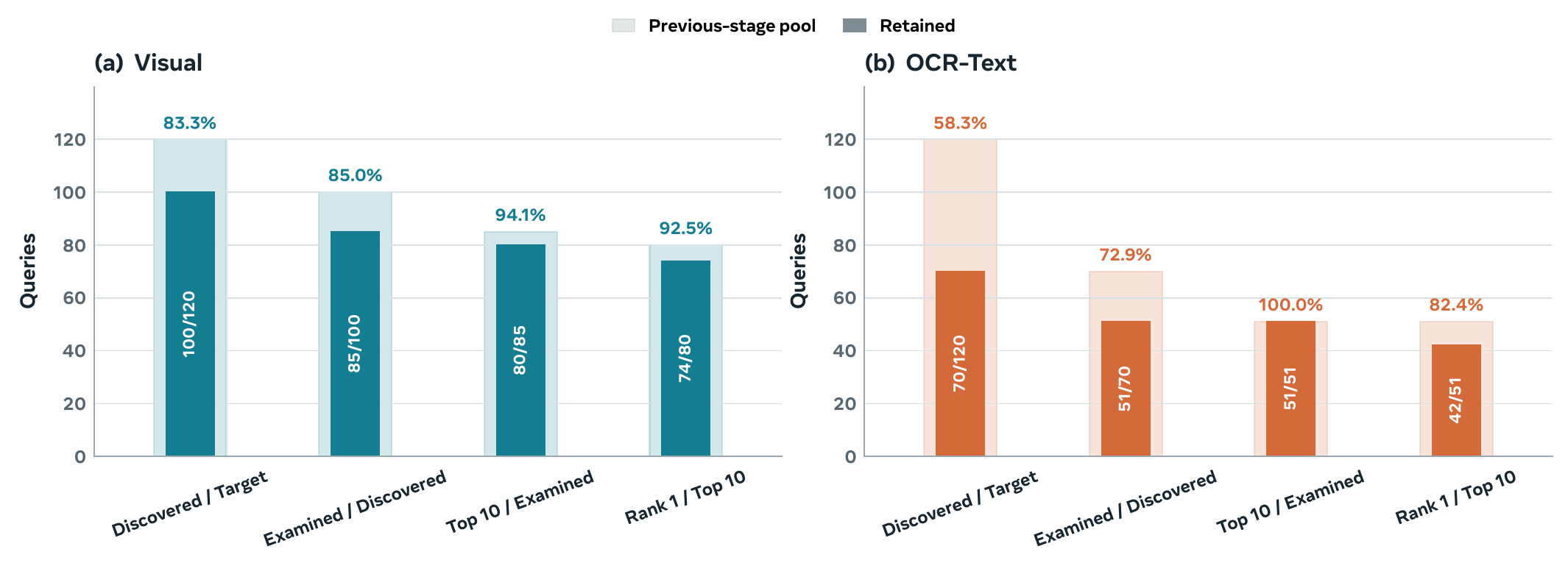}
  \caption{Gold-target progression through the full GPT-5.6-sol traces for
  (a) Visual and (b) OCR-Text retrieval. Pale bars show the previous-stage
  pool, while solid bars show the retained subset. Each percentage reports
  retention from the corresponding pool.}
  \label{fig:target-progression}
\end{figure}

We use GPT-5.6-sol to test how complete support context changes retrieval and
to locate the remaining bottlenecks.  Table~\ref{tab:support-intervention}
compares the standard agent with the support-provided condition, whose
experimental setup is described in Section~\ref{sec:baselines-implementation}.
Figure~\ref{fig:target-progression}
tracks target discovery, examination, and final ordering.  Appendix~\ref{app:target-progression}
defines the target-progression events used in Figure~\ref{fig:target-progression}
and reports their level-wise breakdown in
Figure~\ref{fig:target-progression-levels}.

\begin{table}[htbp]
  \centering
  \small
  \setlength{\tabcolsep}{3.8pt}
  \caption{GPT-5.6-sol retrieval with standard and support-provided initialization on L2 and L3 queries (\%).}
  \label{tab:support-intervention}
  \begin{tabular}{llrrrrr}
    \toprule
    \textbf{Level} & \textbf{Condition} & \textbf{R@1} & \textbf{R@3} & \textbf{R@5} & \textbf{R@10} & \textbf{MRR@10} \\
    \midrule
    \multicolumn{7}{l}{\textbf{Visual}} \\
    \midrule
    \multirow{2}{*}{L2} & Standard & 60.00 & 62.50 & 62.50 & 62.50 & 61.25 \\
      & Support provided & 72.50 & 72.50 & 72.50 & 75.00 & 72.86 \\
    \cmidrule{1-7}
    \multirow{2}{*}{L3} & Standard & 40.00 & 45.00 & 47.50 & 52.50 & 43.36 \\
      & Support provided & 45.00 & 62.50 & 65.00 & 75.00 & 55.19 \\
    \midrule
    \multicolumn{7}{l}{\textbf{OCR-Text}} \\
    \midrule
    \multirow{2}{*}{L2} & Standard & 40.00 & 40.00 & 40.00 & 40.00 & 40.00 \\
      & Support provided & 52.50 & 52.50 & 52.50 & 52.50 & 52.50 \\
    \cmidrule{1-7}
    \multirow{2}{*}{L3} & Standard & 27.50 & 32.50 & 40.00 & 45.00 & 32.12 \\
      & Support provided & 37.50 & 45.00 & 47.50 & 52.50 & 41.40 \\
    \bottomrule
  \end{tabular}
\end{table}

\paragraph{Complete support context strengthens evidence-conditioned retrieval.}
Providing complete support context improves every reported metric across both
retrieval routes and both path depths in Table~\ref{tab:support-intervention}.
MRR@10 rises by 9.28--12.50 points across the four comparisons, showing that
the benefit extends across retrieval representations and query levels.  The
largest effect appears on visual L3 coverage, where R@10 improves by 22.50
points.  Complete support context therefore helps the planner connect evidence
across longer paths, bringing more targets into the final ranking and improving
their ordering.

\paragraph{Target progression from discovery to ranking.}
Figure~\ref{fig:target-progression} localizes the visual--OCR-text gap mainly to
target discovery and examination.
Visual search discovers 83.3\% of targets, compared with 58.3\% for OCR-text, and
agents examine 85.0\% of visual discoveries versus 72.9\% of OCR-text
discoveries.  Since both routes share the planner and visual-inspection tools,
this divergence indicates that visual search provides a stronger corpus-wide
discovery interface.  Once examined, at least 94.1\% of targets enter the final
top 10 and at least 82.4\% of those retained are ranked first.  Final reranking
is therefore comparatively reliable after target inspection.

\paragraph{Qualitative trace examples.}
Appendix~\ref{app:qualitative-trajectories} contrasts a successful and a failed
L3 visual-route episode in Figure~\ref{fig:appendix-traces} and
Tables~\ref{tab:trace-success}--\ref{tab:trace-failure}.  The success uses an
examined support page to reformulate the search and surface the target.  The
failed trace remains anchored to the paper containing the initial brightness
example and never follows the benchmark contrast into the target document.
The second support and target surface only in its final search, remain
unexamined, and the target is omitted from the submitted top 10.  Together,
the examples show how evidence acquisition can redirect retrieval or fail when
the planner does not preserve a cross-document relation.

\section{Conclusion}
\label{sec:conclusion}

We introduced \benchmark{}, a closed-corpus benchmark for visually rich
document retrieval under static and agentic paradigms, constructed from
relation-preserving evidence paths and full-document audits.  A strong
late-interaction retriever is nearly perfect on direct items yet rarely solves
two-bridge queries.  Controlled ablations show that iterative search improves
retrieval, but wide planner variation shows that tool access alone does not
yield reliable retrieval.  Visual search is more effective for corpus-wide
discovery, while page inspection and targeted text access provide complementary
verification.  Providing complete support context further improves ranking on
the same queries.  Trace analysis localizes remaining failures to target
discovery, candidate examination, and evidence-role integration.  These
findings motivate retrieval agents that combine modality-preserving
representations, evidence-directed policies, and role-aware state.

\section*{Limitations}

The current 120-query benchmark is constructed from English-language scientific papers. This controlled domain supports reproducible rendering and full-document verification, but recurring academic layouts and terminology may limit transfer to forms, manuals, slides, interface captures, and multilingual documents. Its fixed corpus and page-level targets also leave collection updates and finer retrieval granularities untested. Moreover, the small topic and cross-document subsets allow only descriptive analysis of those settings. These choices enable rigorous path and hard-negative auditing while narrowing the scope of our claims. Future work can extend the same construction protocol across document genres, languages, corpus scales, and target granularities.

\bibliography{references}
\bibliographystyle{assets/plainnat}

\clearpage
\appendix
\section{Benchmark Specification}
\label{app:benchmark-schema}

\subsection{Task Instantiation}
\label{app:task-instantiation}

In \benchmark{}, a visual document unit $x_i$ corresponds to one rendered page
and is addressed through an opaque handle during evaluation.  For a query $q$,
$a_q$ is its single adjudicated target page.  The reviewed relation-preserving
path determines which non-target pages instantiate $E_q$ and records how their
evidence connects the query to $a_q$.  Thus, $E_q$ is empty for direct items
and contains the intermediate evidence pages for bridged items.

Under this instantiation, $\operatorname{Rel}(q,a_q\mid E_q)=1$ means that
$a_q$ satisfies the query's complete semantic, relational, and visual
conditions when the recorded corpus context is considered.  Plausible
alternatives are checked against the complete request through the hard-negative
and false-negative validation protocol in
Appendix~\ref{app:hard-negative-validation}.

\subsection{Evidence-Path Structure and Query Constraints}
\label{app:path-query-constraints}

For construction, $A$ denotes a query anchor, $B_i$ a semantic bridge, $V$ a
visual target descriptor, $r_i$ a directed relation grounded within one
visual unit, and $\approx$ a verified semantic alignment between semantic
bridges from different units.  The three
templates are
\begin{equation}
  \begin{cases}
    \mathrm{L1:} & A\xrightarrow{r_1}V, \\
    \mathrm{L2:} & A\xrightarrow{r_1}B_1\approx B_2
    \xrightarrow{r_2}V, \\
    \mathrm{L3:} & A\xrightarrow{r_1}B_1\approx B_2
    \xrightarrow{r_2}B_3\approx B_4\xrightarrow{r_3}V.
  \end{cases}
\end{equation}
A semantic alignment may connect two units in one source document or, under the
stricter named-entity rule described in Appendix~\ref{app:automatic-construction}, two
documents.  The number of relation-bearing evidence units defines the level and
its path template.

\begin{table}[htbp]
  \centering
  \small
  \setlength{\tabcolsep}{4pt}
  \caption{Query exposure and hiding policy. Generic referents preserve the
  directed relation skeleton while hiding the semantic bridge identities used
  to construct the item.}
  \label{tab:exposure-policy}
  \begin{tabular}{p{0.06\textwidth}
                  p{0.18\textwidth}
                  p{0.22\textwidth}
                  p{0.46\textwidth}}
    \toprule
    \textbf{Level} & \textbf{Required exposure} & \textbf{Hidden content} & \textbf{Example} \\
    \midrule
    L1 & $A,r_1,V$ & identifiers and captions &
      ``Find the unit where an error flag is shown as a red row in a comparison matrix.'' \\
    \midrule
    L2 & $A,r_1,r_2,V$ & $B_1,B_2,\approx$ &
      ``Find the unit where a policy instantiates a mechanism that is summarized by a stacked latency chart.'' \\
    \midrule
    L3 & $A,r_1,r_2,r_3,V$ & $B_1,\ldots,B_4$ and both $\approx$ links &
      ``Find the unit where a condition produces an intermediate setting, whose evaluation yields a result displayed in a grouped plot.'' \\
    \bottomrule
  \end{tabular}
\end{table}

Table~\ref{tab:exposure-policy} states the query contract. All directed
relations are expressed in natural language, but semantic bridge identities
and semantic alignments are replaced by generic referents. Across all levels,
the query must request a visual unit while withholding paper
titles, page numbers, figure or table numbers, exact captions, target
identifiers, and construction terminology.  The final visual phrase must
support verification while preserving the necessity of the semantic
constraints.

\subsection{Annotation Roles and Inference Boundary}
\label{app:annotation-fields}

\begin{table}[htbp]
  \centering
  \small
  \setlength{\tabcolsep}{4pt}
  \caption{Annotation roles and their inference-time visibility.}
  \label{tab:annotation-fields}
  \begin{tabular}{p{0.20\textwidth}p{0.29\textwidth}p{0.43\textwidth}}
    \toprule
    \textbf{Role} & \textbf{Annotation content} & \textbf{Visibility and use} \\
    \midrule
    Agent-visible input & Final natural-language query; corpus pages exposed through opaque handles &
      The planner sees the query text and handles returned by tools. Handles are randomly assigned with a frozen seed and contain no document or page-order signal. \\
    \midrule
    Evaluation metadata & Evidence-path level, topic, path scope, and review outcomes &
      Available to evaluators for stratification, audit, and data analysis only. \\
    \midrule
    Target and construction annotations & Adjudicated target page; role-specific descriptors and directed relations; semantic alignments; evidence-path pages; query-exposure constraints &
      Supports scoring and construction audits while remaining strictly withheld from the agent. \\
    \bottomrule
  \end{tabular}
\end{table}

Table~\ref{tab:annotation-fields} separates agent-visible input from evaluation
metadata and construction annotations. A standard episode supplies only the
final query and an opaque handle space over the corpus images. Evidence-path
level and topic are reserved for post-prediction stratification and omitted from
the planner prompt.

\section{Corpus and Construction Details}
\label{app:construction}

\subsection{Topic and Document Selection}
\label{app:topic-selection}

We queried the arXiv API in the first half of 2026 with two or three explicit
Boolean queries for each of ten topics.  The search returned 1,722 topic-query
records before within-topic deduplication.  We retained papers published during
this period and scored title, abstract, and category metadata using
topic-specific required, preferred, and excluded phrases.  We selected the ten
highest-scoring globally unique papers per topic.  A required-phrase match
contributed 8 points, a preferred phrase 3, an excluded phrase $-12$, the year
condition 10, and a computer-science/statistics/electrical-engineering
primary category 2.

The topic abbreviations used below are: T1 embodied world models; T2 GUI and
computer-use agents; T3 software-engineering agents; T4 agent memory and
long-horizon context; T5 multimodal RAG and visual documents; T6 test-time
reasoning; T7 controllable video generation; T8 3D and spatial intelligence;
T9 efficient inference and KV-cache optimization; and T10 scientific
foundation models.

\begin{table}[htbp]
  \centering
  \scriptsize
  \setlength{\tabcolsep}{1.8pt}
  \caption{Complete source-document inventory. ``T,'' ``Cat.,'' and ``Pg.''
  denote topic identifier, primary arXiv category, and rendered page count in
  the 2,375-unit retrieval corpus, respectively.}
  \label{tab:document-inventory}
  \resizebox{\linewidth}{!}{%
  \begin{tabular}{lclr|lclr|lclr|lclr}
    \toprule
    \textbf{arXiv} & \textbf{T} & \textbf{Cat.} & \textbf{Pg.} &
    \textbf{arXiv} & \textbf{T} & \textbf{Cat.} & \textbf{Pg.} &
    \textbf{arXiv} & \textbf{T} & \textbf{Cat.} & \textbf{Pg.} &
    \textbf{arXiv} & \textbf{T} & \textbf{Cat.} & \textbf{Pg.} \\
    \midrule
    2601.18203 & T5 & cs.IR & 12 & 2605.07467 & T10 & cs.LG & 17 & 2606.05773 & T1 & cs.RO & 14 & 2606.18478 & T7 & cs.CV & 34 \\
    2602.06090 & T3 & cs.SE & 16 & 2605.13986 & T6 & cs.LG & 83 & 2606.06322 & T2 & cs.AI & 19 & 2606.18847 & T4 & cs.AI & 27 \\
    2602.06593 & T3 & cs.SE & 21 & 2605.17641 & T4 & cs.AI & 12 & 2606.06891 & T8 & cs.CV & 32 & 2606.19340 & T8 & cs.RO & 21 \\
    2602.19961 & T5 & cs.CL & 35 & 2605.17965 & T3 & cs.SE & 45 & 2606.06915 & T6 & cs.CL & 15 & 2606.19531 & T1 & cs.CV & 19 \\
    2602.20502 & T2 & cs.AI & 12 & 2605.17976 & T10 & cs.AI & 31 & 2606.08688 & T10 & cs.RO & 20 & 2606.19746 & T9 & cs.DC & 22 \\
    2602.22013 & T5 & cs.CV & 10 & 2605.17985 & T10 & cs.LG & 22 & 2606.08737 & T1 & cs.RO & 16 & 2606.20512 & T3 & cs.SE & 20 \\
    2602.24134 & T5 & cs.CV & 26 & 2605.19943 & T6 & cs.AI & 14 & 2606.09864 & T9 & cs.LG & 61 & 2606.20729 & T10 & physics.chem-ph & 17 \\
    2603.01666 & T5 & cs.CL & 27 & 2605.21312 & T9 & cs.DC & 20 & 2606.10183 & T7 & cs.CV & 10 & 2606.20891 & T7 & cs.CV & 29 \\
    2603.07432 & T2 & cs.CV & 25 & 2605.22829 & T5 & cs.IR & 19 & 2606.10662 & T6 & cs.MA & 23 & 2606.21023 & T9 & cs.LG & 25 \\
    2603.08013 & T2 & cs.AI & 16 & 2605.23057 & T9 & cs.LG & 11 & 2606.11078 & T2 & cs.AI & 29 & 2606.21633 & T9 & cs.LG & 14 \\
    2603.08533 & T2 & cs.CV & 16 & 2605.23163 & T9 & cs.CL & 17 & 2606.11576 & T6 & cs.CV & 30 & 2606.22082 & T3 & cs.SE & 36 \\
    2603.12823 & T2 & cs.CL & 16 & 2605.27570 & T6 & cs.AI & 22 & 2606.13035 & T7 & cs.CV & 17 & 2606.22617 & T8 & cs.CV & 14 \\
    2603.15921 & T3 & cs.SE & 17 & 2605.29074 & T8 & cs.CV & 11 & 2606.13494 & T1 & cs.RO & 22 & 2606.22631 & T8 & cs.CV & 28 \\
    2603.16085 & T8 & cs.CV & 25 & 2605.29639 & T9 & cs.OS & 14 & 2606.13672 & T1 & cs.RO & 32 & 2606.22694 & T8 & cs.CV & 28 \\
    2603.17826 & T3 & cs.SE & 15 & 2605.30917 & T5 & cs.IR & 12 & 2606.13740 & T9 & cs.LG & 14 & 2606.22813 & T6 & cs.AI & 53 \\
    2603.26211 & T2 & cs.CV & 14 & 2606.00756 & T4 & cs.AI & 19 & 2606.13768 & T7 & cs.CV & 32 & 2606.22844 & T4 & cs.AI & 36 \\
    2604.05012 & T9 & cs.AR & 8 & 2606.00832 & T4 & cs.CL & 9 & 2606.14162 & T7 & cs.CV & 17 & 2606.22906 & T3 & cs.SE & 12 \\
    2604.14609 & T10 & cs.AI & 65 & 2606.01613 & T5 & cs.IR & 24 & 2606.14870 & T10 & hep-ph & 33 & 2606.22938 & T6 & cs.LG & 38 \\
    2604.15821 & T10 & cs.DC & 11 & 2606.02355 & T4 & cs.AI & 15 & 2606.14948 & T3 & cs.SE & 19 & 2606.23296 & T1 & cs.RO & 16 \\
    2604.17288 & T3 & cs.AR & 10 & 2606.02800 & T1 & cs.CV & 139 & 2606.15001 & T10 & physics.comp-ph & 15 & 2606.23557 & T8 & cs.CV & 41 \\
    2604.23276 & T5 & cs.CV & 8 & 2606.03099 & T4 & cs.CL & 17 & 2606.15032 & T1 & cs.LG & 24 & 2606.23610 & T7 & cs.CV & 24 \\
    2604.23941 & T2 & cs.CV & 20 & 2606.04432 & T7 & cs.CV & 13 & 2606.17257 & T7 & cs.CV & 17 & 2606.24480 & T10 & cond-mat.mtrl-sci & 54 \\
    2604.27253 & T2 & cs.AI & 21 & 2606.04527 & T7 & cs.MM & 21 & 2606.17803 & T6 & cs.LG & 17 & 2606.24626 & T4 & cs.AI & 16 \\
    2604.27996 & T4 & cs.AI & 5 & 2606.05145 & T6 & cs.LG & 43 & 2606.18180 & T1 & cs.CV & 16 & 2606.26800 & T8 & cs.RO & 8 \\
    2605.06460 & T5 & cs.LG & 21 & 2606.05761 & T4 & cs.AI & 47 & 2606.18375 & T1 & cs.RO & 20 & 2606.27876 & T8 & cs.CV & 10 \\
    \bottomrule
  \end{tabular}}
\end{table}

Table~\ref{tab:document-inventory} lists every selected document, primary arXiv
category, and rendered page count. Licensing is source-version specific across
the 100 documents; the applicable terms are those attached to each versioned
arXiv record.

\subsection{Automated Descriptor Extraction, Semantic Alignment, and Path Validation}
\label{app:automatic-construction}

PDFs were rendered at 144 dpi and their text layers extracted with PyMuPDF.
Pages with fewer than 10 extracted words or a reference/boilerplate heading
remained in the retrieval corpus and were excluded from query construction.  A
GPT-5.5 VLM at medium reasoning effort and temperature 0 then received each
page image, page text, and document context.  Its structured prompt requested
up to three query anchors, five semantic bridges, three visual target
descriptors, and eight typed directed relations.  The validation pass enforced
role compatibility, query-anchor non-revelation, semantic-bridge stability,
visual distinctiveness of target descriptors, and grounding of directed relations.
The extraction pass produced 21,955 descriptors and 17,693 directed relations.
Validation retained 20,251 descriptors---6,149 query anchors, 10,531 semantic
bridges, and 3,571 visual target descriptors---together with 15,390 usable
directed relations.

Qwen3-Embedding-8B embedded 10,531 usable semantic bridges.  Matching was
performed independently within each topic with cosine threshold 0.72, ten
neighbors per semantic bridge, and a cap of 1,000 pairs per topic.  GPT-5.5
then judged
whether each pair denoted the same or a contextually compatible concept and
whether the overlap was too generic.  Same-document semantic alignments permit
stable mid-level concepts; cross-document semantic alignments must connect a
proper name, alias, acronym, or variant of the same named entity.  Path
composition retained only verified semantic alignments and used per-topic caps
of 120 L1, 200 L2, and 200 L3 paths.

The final automatic check used level-specific JSON schemas.  L1 required a
meaningful directed relation and visually necessary target.  L2/L3 additionally
tested multihop coherence, semantic-bridge necessity, whether the query anchor
directly identified the target, and whether an earlier unit exposed a later
semantic bridge.  The verbalizability check required every directed relation
and endpoint intended for exposure while forbidding semantic bridge identities,
semantic alignments, titles, page numbers, figure/table identifiers, exact
captions, and path meta-language.  It accepted provisional
wording for 1,501 paths.  A further 21 paths passed the evidence-path checks but
failed wording quality; these remained available for full-document review and
manual final query authoring.

\subsection{Construction Funnel by Topic}
\label{app:topic-funnel-details}

\begin{table}[htbp]
  \centering
  \small
  \setlength{\tabcolsep}{5pt}
  \caption{Construction funnel by topic. ``Descr.,'' ``Rel.,'' and ``Cand.''
  denote validated descriptors, usable directed relations, and candidate
  semantic-bridge pairs, respectively. ``Verified,'' ``Paths,'' ``Valid,'' and
  ``Final'' count verified alignments, composed paths, automatically validated
  paths, and selected benchmark items.}
  \label{tab:topic-funnel}
  \begin{tabular}{lrrrrrrr}
    \toprule
    \textbf{Topic} & \textbf{Descr.} & \textbf{Rel.} & \textbf{Cand.} & \textbf{Verified} & \textbf{Paths} & \textbf{Valid} & \textbf{Final} \\
    \midrule
    T1 & 2,700 & 2,049 & 1,000 & 426 & 329 & 184 & 13 \\
    T2 & 1,570 & 1,192 & 1,000 & 358 & 249 & 114 & 10 \\
    T3 & 1,782 & 1,333 & 1,000 & 487 & 289 & 148 & 19 \\
    T4 & 1,750 & 1,347 & 1,000 & 322 & 201 & 107 & 8 \\
    T5 & 1,618 & 1,154 & 1,000 & 354 & 232 & 115 & 11 \\
    T6 & 2,901 & 2,236 & 1,000 & 576 & 385 & 151 & 13 \\
    T7 & 1,848 & 1,432 & 1,000 & 486 & 462 & 220 & 10 \\
    T8 & 1,855 & 1,425 & 1,000 & 481 & 320 & 178 & 13 \\
    T9 & 1,780 & 1,372 & 1,000 & 581 & 450 & 153 & 12 \\
    T10 & 2,447 & 1,850 & 1,000 & 637 & 411 & 152 & 11 \\
    \midrule
    Total & 20,251 & 15,390 & 10,000 & 4,708 & 3,328 & 1,522 & 120 \\
    \bottomrule
  \end{tabular}
\end{table}

Table~\ref{tab:topic-funnel} reports the complete topic-local funnel. Before
automatic validation, the 3,328 composed paths comprise 425 L1, 1,806
L2, and 1,097 L3 paths, including 425 single-unit, 2,746 same-document, and 157
cross-document paths.  The 1,522 surviving paths comprise 364 L1, 1,058 L2,
and 100 L3 paths.

The fixed 1,000-pair cap gives each topic the same semantic-alignment
verification budget.  Subsequent counts vary after verification and path
validation because topics differ in the availability of specific semantic
alignments, visually structured target pages, and shortcut-free L3 paths.
Final selection prioritizes reviewed path quality; among paths of comparable
quality, it favors broader topic coverage.  Cross-document paths are rarer
because they require named-entity alignments and full-context validation of all
involved documents.

\section{Audit and Curation Protocol}
\label{app:audit}

\subsection{Full-Document Path Review and Scoring}
\label{app:path-review}

All 1,522 automatically validated paths were reviewed, including paths whose
provisional wording failed the automatic language check.  For a same-document
path, the complete source document was read; for a cross-document path, every
participating document was read.  Ninety-six of the 100 source documents
appear in this queue.  The other four contribute no automatically validated
path and therefore require no path-level judgment.

One author and a separate AI reviewer independently scored
each path on a 0--100 scale in full-document context.  Both considered semantic
validity, necessity of the intermediate evidence, semantic-bridge specificity,
visual grounding, whether a natural non-leaking query could be written, and diversity
relative to the pool.  The two scores were produced in separate passes.  A
path remained eligible only when both scores were at
least 60; eligible paths were then ordered by their arithmetic mean.  This
double gate prevents one high score from masking a severe objection raised by
the other review.

\subsection{Selection and Target Deduplication}
\label{app:path-selection}

Selection was path-quality first.  Within each level, topic, source-document,
visual-form, and reasoning-pattern balance distinguish paths of comparable
quality.  When two queries point to the same target image, the higher
evidence-path level is preferred, followed by the higher mean score within a
level.  Replacements come from the complete eligible pool and receive the same
full-document check.  The final set contains 40 queries at each level and 120
distinct target images.  Its scope distribution is 40 single-unit, 74
same-document, and 6 cross-document paths, spanning 55 documents across their
recorded evidence paths.

\subsection{Constraint-Guided Final Query Authoring}
\label{app:query-authoring}

Final query writing began only after path selection.  The author used the
selected path and full-document context to (i) expose the query anchor and every
directed relation in the level template, (ii) replace semantic bridge identities
and semantic alignments with coherent generic referents, (iii) retain a visual
target descriptor while omitting captions and unique identifiers,
and (iv) formulate the request at visual-unit granularity.  For L2
and L3, omitting a relation invalidates the wording even when the endpoints
remain recognizable.  For L1, the direct relation must likewise remain
explicit.  Automatic verbalizations served only as expressibility checks.

Visual granularity was calibrated against the full corpus.  A phrase was
broadened when it encoded a near-pixel-level fingerprint, and narrowed only by
the minimal observable distinction needed to exclude a verified alternative.
The resulting 120 queries average 25.6 whitespace-delimited words, with a
median of 26 and a range of 15--39.

\subsection{Hard-Negative and False-Negative Validation}
\label{app:hard-negative-validation}

After reading each involved source paper in full, the reviewer checked all of
its pages for plausible alternatives satisfying the complete query.
Qwen3-VL-Embedding-8B additionally retrieved ten page images from the full
corpus to surface high-similarity alternatives beyond this document-wide
review.  Every non-gold result was inspected, regardless of whether the gold
target was already retrieved.  Layout, object, or phrase overlap alone was
insufficient; a valid candidate had to satisfy all constraints expressed by the
query anchor, directed relations, and visual target descriptor.

When this validation identifies a complete alternative, the reviewer adds the
minimum observable distinction between the target and that alternative or
replaces the item.  The resulting query is rechecked against the retrieved
candidates and the relevant full-document context.  These checks concentrate
manual review on high-risk alternatives surfaced by a strong retriever.

\section{Baseline Protocols}
\label{app:baseline-protocols}

\subsection{Embedding Indices}
\label{app:embedding-indices}

The visual single-vector index contains one normalized Qwen3-VL-Embedding-8B vector for each
of the 2,375 rendered page images.  The OCR-text index first parses every page
image with PaddleOCR-VL-1.6 using PP-DocLayoutV3, concatenates the recovered
plain-text blocks in reading order, and embeds that text with
Qwen3-Embedding-8B.  Static retrieval encodes the benchmark query, whereas an
agent search encodes the planner-issued search string; neither route adds a
retrieval instruction.  Cosine similarity
over normalized vectors determines the returned order.  The page-image
vectors, OCR records, and OCR-text vectors are cached once and reused by static
and agent baselines.

The sparse OCR baseline uses Okapi BM25 with $k_1=1.2$ and $b=0.75$.  The
hybrid OCR baseline takes the top 1,000 BM25 and dense results and combines
their ranks with reciprocal-rank fusion using $k=60$.
Nemotron-ColEmbed-VL-8B-V2 encodes each query and page as token-level vectors and scores them with the
model's MaxSim operator.  It uses the model-provided processor, BF16 weights,
and no additional query instruction.

\subsection{Agent Tools}
\label{app:agent-tools}

\begin{table}[htbp]
  \centering
  \footnotesize
  \setlength{\tabcolsep}{3.5pt}
  \caption{Frozen information-tool and terminal-action schemas.  The two
  global search tools are mutually exclusive: no evaluated agent receives a
  hybrid index.}
  \label{tab:tool-schemas}
  \begin{tabular}{llp{0.29\textwidth}p{0.32\textwidth}}
    \toprule
    \textbf{Route} & \textbf{Action} & \textbf{Arguments} & \textbf{Observation and constraints} \\
    \midrule
    \multirow[t]{2}{*}{Visual} & \texttt{visual\_search} & \texttt{query}: string; \texttt{top\_k}: integer $[1,50]$ & Ordered \texttt{page\_handle}, score pairs from the page-image index. \\
      & \texttt{ocr\_image} & \texttt{target\_handle}: one searched full page or generated crop & Plain OCR text and blocks from PaddleOCR-VL-1.6; accepts one image per action. \\
    \midrule
    \multirow[t]{2}{*}{OCR-Text} & \texttt{text\_search} & \texttt{query}: string; \texttt{top\_k}: integer $[1,50]$ & Ordered \texttt{page\_handle}, score pairs from the OCR-text index; no snippets. \\
      & \texttt{get\_page\_ocr} & \texttt{page\_handle}: one searched full page & Complete cached plain OCR text; no OCR model is rerun and crop handles are rejected. \\
    \midrule
    \multirow[t]{3}{*}{Both} & \texttt{inspect\_pages} & \texttt{page\_handles}: 1--10 distinct searched pages & Full-page image artifacts plus handle and image dimensions.  The call is atomic. \\
      & \texttt{crop\_pages} & \texttt{crops}: 1--10 objects containing \texttt{page\_handle}, normalized $[x_1,y_1,x_2,y_2]$, and optional \texttt{purpose} & Cropped image artifacts and crop handles.  The source page must have been inspected. Coordinates use top-left $(0,0)$ and bottom-right $(1,1)$. Valid entries survive other entries' failures; no maximum area is imposed. \\
      & \texttt{submit\_answer} & Up to ten distinct ranked objects, each with \texttt{page\_handle}, \texttt{evidence\_page\_handles}, and a short rationale; one overall rationale & Terminates the episode.  Exactly ten are required when available. Only discovered pages may be ranked; evidence handles must have been read or viewed. \\
    \bottomrule
  \end{tabular}
\end{table}

Table~\ref{tab:tool-schemas} gives the complete inference interface.  Search
results expose only opaque handles and scores, withholding snippets, source
titles, document identifiers, and page numbers.  Image-returning tools
deliver image artifacts to the next planner turn in addition to the compact
JSON status shown in the table.  Full-page inspection, regional cropping, and
on-demand text recognition follow recent visual-agent designs
\citep{he2026vistahop,wang2026agenticocr}.

\subsection{Planner and Episode Configuration}
\label{app:planner-configuration}

The closed-source planners are GPT-5.5, GPT-5.6-luna, GPT-5.6-terra,
GPT-5.6-sol, Claude Opus 4.8, Claude Fable 5, Claude Sonnet 5, and Claude Opus
5.  We also evaluate the open-weight Qwen3.5-397B-A17B with thinking enabled
and disabled.  Reasoning effort is set to \texttt{medium} where supported.

The neutral planner prompt states the closed-corpus objective, enumerates the
available tools, explains that handles are opaque, and requires exactly one
JSON action per turn.  It omits the gold path, evidence units, document
identity, topic, and level.  The planner may choose an information action or
submit a ranking during each step of a 12-step interaction budget.  The ranking
must contain ten distinct pages whenever ten candidates have been discovered,
and rank-one accuracy remains the primary objective.  Per-rank rationales and
evidence handles are logged for audit and excluded from scoring.

If no valid ranking has been submitted when the 12-step interaction budget is
exhausted, the harness makes one ranking-only finalization call
that exposes the eligible discovered and examined handles but disables all
information tools.  A malformed evidence list is normalized by removing
unseen evidence handles while preserving the ranking.  Duplicate, unknown, or
wrong-length rankings remain invalid.

\subsection{Ablation Configurations}
\label{app:ablation-configurations}

All ablations use GPT-5.6-sol and retain the remaining route settings.  The
no-iterative-search variant fixes its candidate set to the 30 highest-ranked
pages from the corresponding Qwen3 retriever, judges ten independent three-page
batches from full page images and cached OCR, and aggregates the judgments into
a top-10 ranking with one final call.  This candidate-set size slightly exceeds
the full agent's mean page exposure on both routes and provides a common
verification budget.  The fixed pipeline cannot reformulate searches or
replace candidates.  The other variants remove regional
cropping, route-specific text access, or full-page inspection; removing page
inspection also disables cropping, which depends on an inspected source page.

\subsection{Inference Resources and Infrastructure}
\label{app:runtime-accounting}

\begin{table}[htbp]
  \centering
  \small
  \setlength{\tabcolsep}{4pt}
  \caption{Per-query resource indicators for valid GPT-5.6-sol episodes. Input
  and output token counts sum usage over all model calls in an episode.}
  \label{tab:gpt56-resource}
  \begin{tabular}{llrrrr}
    \toprule
    \textbf{System} & \textbf{Route} & \textbf{Model calls} & \textbf{Pages viewed} & \textbf{Input tokens} & \textbf{Output tokens} \\
    \midrule
    No iterative search & Visual & 11.00 & 30.00 & 101,075 & 8,932 \\
    No iterative search & OCR-Text & 11.00 & 30.00 & 124,925 & 8,597 \\
    Full agent & Visual & 8.34 & 25.22 & 177,056 & 2,250 \\
    Full agent & OCR-Text & 10.18 & 28.22 & 248,125 & 2,632 \\
    \bottomrule
  \end{tabular}
\end{table}

Table~\ref{tab:gpt56-resource} compares the measured per-query inference
resources of the no-iterative-search control and the full agent.  With 30 page
observations and 11 model calls, the fixed pipeline matches or exceeds the
full-agent averages on both routes.  The resulting page exposure and model-call
counts place the two systems on a comparable scale, focusing the comparison on
whether intermediate observations can redirect subsequent retrieval.  Their
token profiles reflect the two execution structures: independent batch judgments
produce more output tokens, whereas the accumulating full-agent history
produces more input tokens.

Page rendering uses PyMuPDF at 144 dpi.  OCR records and embedding indices are
cached before evaluation.  The visual and OCR-text routes run on separate
NVIDIA A100-SXM4-80GB devices.  Immutable caches are shared read-only, while
runtime crops and episode state remain isolated.

\section{Detailed Results}
\label{app:detailed-results}

\subsection{Level-Wise Main Results}
\label{app:level-main-results}

\begin{table}[htbp]
  \centering
  \footnotesize
  \setlength{\tabcolsep}{2.0pt}
  \caption{Level-wise retrieval results on \benchmark{} (\%). Each level contains 40 queries. Invalid completed episodes are scored as zero. Bold marks the best result for each route, level, and metric; tied best values are all bold.}
  \label{tab:level-results}
  \begin{tabular}{llrrrrrrrrrr}
    \toprule
    \multirow{2}{*}{\textbf{Retriever / Planner}} & \multirow{2}{*}{\textbf{Level}} &
    \multicolumn{5}{c}{\textbf{Visual}} & \multicolumn{5}{c}{\textbf{OCR-Text}} \\
    \cmidrule(lr){3-7}\cmidrule(lr){8-12}
      & & \textbf{R@1} & \textbf{R@3} & \textbf{R@5} & \textbf{R@10} & \textbf{MRR@10}
      & \textbf{R@1} & \textbf{R@3} & \textbf{R@5} & \textbf{R@10} & \textbf{MRR@10} \\
    \midrule
    \multicolumn{12}{l}{\textbf{Static Retrievers}} \\
    \midrule
    \multirow{3}{*}{Qwen3 Embedding}
      & L1 & 40.00 & 52.50 & 60.00 & 72.50 & 49.11 & 2.50 & 5.00 & 10.00 & 17.50 & 5.84 \\
      & L2 & 20.00 & 27.50 & 30.00 & 40.00 & 25.30 & 2.50 & 10.00 & 15.00 & 20.00 & 7.50 \\
      & L3 & 2.50 & 10.00 & 12.50 & 22.50 & 7.62 & 0.00 & 2.50 & 15.00 & 20.00 & 4.68 \\
    \midrule
    \multirow{3}{*}{BM25}
      & L1 & -- & -- & -- & -- & -- & 15.00 & 30.00 & 37.50 & 55.00 & 25.59 \\
      & L2 & -- & -- & -- & -- & -- & 5.00 & 12.50 & 15.00 & 22.50 & 9.49 \\
      & L3 & -- & -- & -- & -- & -- & 0.00 & 2.50 & 5.00 & 12.50 & 2.31 \\
    \midrule
    \multirow{3}{*}{BM25 + dense RRF}
      & L1 & -- & -- & -- & -- & -- & 10.00 & 22.50 & 32.50 & 42.50 & 19.55 \\
      & L2 & -- & -- & -- & -- & -- & 5.00 & 12.50 & 17.50 & 22.50 & 10.40 \\
      & L3 & -- & -- & -- & -- & -- & 0.00 & 7.50 & 7.50 & 15.00 & 4.28 \\
    \midrule
    \multirow{3}{*}{Nemotron ColEmbed}
      & L1 & \textbf{97.50} & \textbf{97.50} & \textbf{100.00} & \textbf{100.00} & \textbf{98.00} & -- & -- & -- & -- & -- \\
      & L2 & 20.00 & 42.50 & 65.00 & 70.00 & 36.34 & -- & -- & -- & -- & -- \\
      & L3 & 2.50 & 17.50 & 30.00 & 40.00 & 12.24 & -- & -- & -- & -- & -- \\
    \midrule
    \multicolumn{12}{l}{\textbf{Tool-Using Agents: Closed-Source Planners}} \\
    \midrule
    \multirow{3}{*}{GPT-5.5}
      & L1 & 75.00 & 75.00 & 77.50 & 80.00 & 75.94 & 30.00 & 30.00 & 30.00 & 32.50 & 30.31 \\
      & L2 & 57.50 & 57.50 & 60.00 & 60.00 & 58.13 & 27.50 & 27.50 & 27.50 & 27.50 & 27.50 \\
      & L3 & 47.50 & 52.50 & 55.00 & 67.50 & 51.56 & 22.50 & 25.00 & 25.00 & 25.00 & 23.75 \\
    \midrule
    \multirow{3}{*}{GPT-5.6-luna}
      & L1 & 65.00 & 65.00 & 65.00 & 72.50 & 66.13 & 20.00 & 20.00 & 20.00 & 27.50 & 21.09 \\
      & L2 & 42.50 & 45.00 & 50.00 & 55.00 & 45.38 & 15.00 & 15.00 & 15.00 & 15.00 & 15.00 \\
      & L3 & 22.50 & 32.50 & 37.50 & 45.00 & 28.75 & 2.50 & 7.50 & 20.00 & 22.50 & 8.23 \\
    \midrule
    \multirow{3}{*}{GPT-5.6-terra}
      & L1 & 77.50 & 80.00 & 80.00 & 82.50 & 79.06 & 32.50 & 32.50 & 32.50 & 37.50 & 33.11 \\
      & L2 & 52.50 & 52.50 & 52.50 & 57.50 & 53.17 & 25.00 & 27.50 & 27.50 & 30.00 & 26.19 \\
      & L3 & 32.50 & 40.00 & 42.50 & 52.50 & 37.73 & 22.50 & 25.00 & 25.00 & 32.50 & 24.46 \\
    \midrule
    \multirow{3}{*}{GPT-5.6-sol}
      & L1 & 85.00 & 87.50 & 87.50 & 90.00 & 86.15 & 42.50 & 42.50 & 42.50 & 47.50 & 43.06 \\
      & L2 & 60.00 & 62.50 & 62.50 & 62.50 & 61.25 & \textbf{40.00} & \textbf{40.00} & 40.00 & 40.00 & \textbf{40.00} \\
      & L3 & 40.00 & 45.00 & 47.50 & 52.50 & 43.36 & 27.50 & 32.50 & 40.00 & 45.00 & 32.12 \\
    \midrule
    \multirow{3}{*}{Claude Opus 4.8}
      & L1 & 70.00 & 72.50 & 75.00 & 82.50 & 72.32 & 22.50 & 22.50 & 25.00 & 40.00 & 24.79 \\
      & L2 & 50.00 & 55.00 & 57.50 & 62.50 & 52.88 & 32.50 & 37.50 & 40.00 & 42.50 & 35.33 \\
      & L3 & 22.50 & 35.00 & 45.00 & 60.00 & 31.97 & 17.50 & 22.50 & 25.00 & 40.00 & 22.01 \\
    \midrule
    \multirow{3}{*}{Claude Fable 5}
      & L1 & 80.00 & 82.50 & 85.00 & 92.50 & 82.57 & 40.00 & 42.50 & 42.50 & \textbf{52.50} & 42.38 \\
      & L2 & 52.50 & 57.50 & 62.50 & 67.50 & 55.88 & 30.00 & 37.50 & 40.00 & \textbf{47.50} & 35.01 \\
      & L3 & \textbf{55.00} & \textbf{62.50} & \textbf{67.50} & \textbf{80.00} & \textbf{61.61} & \textbf{35.00} & \textbf{45.00} & \textbf{47.50} & \textbf{57.50} & \textbf{40.63} \\
    \midrule
    \multirow{3}{*}{Claude Sonnet 5}
      & L1 & 42.50 & 57.50 & 62.50 & 72.50 & 50.75 & 20.00 & 20.00 & 22.50 & 25.00 & 20.94 \\
      & L2 & 32.50 & 42.50 & 42.50 & 52.50 & 38.27 & 15.00 & 15.00 & 20.00 & 20.00 & 16.00 \\
      & L3 & 15.00 & 22.50 & 35.00 & 42.50 & 22.96 & 17.50 & 25.00 & 25.00 & 25.00 & 20.42 \\
    \midrule
    \multirow{3}{*}{Claude Opus 5}
      & L1 & 92.50 & 92.50 & 92.50 & 92.50 & 92.50 & \textbf{47.50} & \textbf{47.50} & \textbf{47.50} & 50.00 & \textbf{47.78} \\
      & L2 & \textbf{62.50} & \textbf{67.50} & \textbf{67.50} & \textbf{72.50} & \textbf{64.69} & 32.50 & \textbf{40.00} & \textbf{42.50} & \textbf{47.50} & 37.05 \\
      & L3 & 47.50 & 52.50 & 55.00 & 60.00 & 51.09 & 32.50 & 37.50 & 45.00 & 50.00 & 36.90 \\
    \midrule
    \multicolumn{12}{l}{\textbf{Tool-Using Agents: Open-Weight Planner}} \\
    \midrule
    \multirow{3}{*}{Qwen3.5 (thinking)}
      & L1 & 47.50 & 62.50 & 67.50 & 67.50 & 55.29 & 15.00 & 15.00 & 17.50 & 22.50 & 16.09 \\
      & L2 & 22.50 & 25.00 & 25.00 & 40.00 & 25.80 & 5.00 & 5.00 & 5.00 & 7.50 & 5.31 \\
      & L3 & 12.50 & 12.50 & 12.50 & 12.50 & 12.50 & 5.00 & 7.50 & 7.50 & 17.50 & 7.69 \\
    \midrule
    \multirow{3}{*}{Qwen3.5 (no thinking)}
      & L1 & 35.00 & 45.00 & 60.00 & 67.50 & 43.71 & 2.50 & 2.50 & 10.00 & 12.50 & 4.56 \\
      & L2 & 17.50 & 17.50 & 20.00 & 32.50 & 19.65 & 2.50 & 2.50 & 5.00 & 5.00 & 3.00 \\
      & L3 & 5.00 & 7.50 & 10.00 & 15.00 & 7.28 & 0.00 & 7.50 & 7.50 & 12.50 & 3.94 \\
    \bottomrule
  \end{tabular}
\end{table}

Table~\ref{tab:level-results} distinguishes direct query--page matching from
retrieval that depends on one or two semantic bridges.  Late interaction is
nearly perfect on direct items yet drops sharply on both bridge levels.
Tool-using agents provide the strongest results on the longer evidence paths.

\subsection{Topic-Level Results}
\label{app:topic-results}

\begin{table}[htbp]
  \centering
  \footnotesize
  \setlength{\tabcolsep}{4pt}
  \caption{GPT-5.6-sol topic-level retrieval results (\%). Topic names are defined in Appendix~\ref{app:topic-selection}.}
  \label{tab:gpt56-topic-results}
  \begin{tabular}{llrrrrrr}
    \toprule
    \textbf{Route} & \textbf{Topic} & \textbf{Queries} & \textbf{R@1} & \textbf{R@3} & \textbf{R@5} & \textbf{R@10} & \textbf{MRR@10} \\
    \midrule
    \multirow{10}{*}{Visual}
      & T1 & 13 & 38.46 & 46.15 & 46.15 & 61.54 & 44.55 \\
      & T2 & 10 & 70.00 & 70.00 & 70.00 & 70.00 & 70.00 \\
      & T3 & 19 & 42.11 & 52.63 & 57.89 & 57.89 & 47.54 \\
      & T4 & 8 & 75.00 & 87.50 & 87.50 & 87.50 & 79.17 \\
      & T5 & 11 & 72.73 & 72.73 & 72.73 & 72.73 & 72.73 \\
      & T6 & 13 & 76.92 & 76.92 & 76.92 & 76.92 & 76.92 \\
      & T7 & 10 & 40.00 & 40.00 & 40.00 & 40.00 & 40.00 \\
      & T8 & 13 & 69.23 & 69.23 & 69.23 & 69.23 & 69.23 \\
      & T9 & 12 & 66.67 & 66.67 & 66.67 & 75.00 & 67.86 \\
      & T10 & 11 & 81.82 & 81.82 & 81.82 & 81.82 & 81.82 \\
    \midrule
    \multirow{10}{*}{OCR-Text}
      & T1 & 13 & 38.46 & 53.85 & 53.85 & 53.85 & 44.87 \\
      & T2 & 10 & 30.00 & 30.00 & 30.00 & 30.00 & 30.00 \\
      & T3 & 19 & 21.05 & 21.05 & 31.58 & 47.37 & 25.75 \\
      & T4 & 8 & 37.50 & 37.50 & 37.50 & 37.50 & 37.50 \\
      & T5 & 11 & 54.55 & 54.55 & 54.55 & 54.55 & 54.55 \\
      & T6 & 13 & 53.85 & 53.85 & 53.85 & 53.85 & 53.85 \\
      & T7 & 10 & 40.00 & 40.00 & 40.00 & 50.00 & 41.00 \\
      & T8 & 13 & 7.69 & 7.69 & 15.38 & 15.38 & 9.62 \\
      & T9 & 12 & 33.33 & 33.33 & 33.33 & 33.33 & 33.33 \\
      & T10 & 11 & 63.64 & 63.64 & 63.64 & 63.64 & 63.64 \\
    \bottomrule
  \end{tabular}
\end{table}

Table~\ref{tab:gpt56-topic-results} reports all five metrics for GPT-5.6-sol by
topic.  The visual route matches or exceeds OCR-text in every topic at R@1.
OCR-text varies more sharply: it reaches 63.64\% R@1 on T10 but only 7.69\% on
T8, where spatial diagrams and view-dependent layouts are poorly represented by
OCR text alone.

\subsection{Level-Wise Ablation Results}
\label{app:level-ablations}

\begin{table}[htbp]
  \centering
  \footnotesize
  \setlength{\tabcolsep}{4pt}
  \caption{Level-wise GPT-5.6-sol ablations on both retrieval routes (\%).}
  \label{tab:level-ablation}
  \begin{tabular}{llrrrrr}
    \toprule
    \textbf{Configuration} & \textbf{Level} & \textbf{R@1} & \textbf{R@3} & \textbf{R@5} & \textbf{R@10} & \textbf{MRR@10} \\
    \midrule
    \multicolumn{7}{l}{\textbf{Visual}} \\
    \midrule
    \multirow{3}{*}{Full agent}
      & L1 & 85.00 & 87.50 & 87.50 & 90.00 & 86.15 \\
      & L2 & 60.00 & 62.50 & 62.50 & 62.50 & 61.25 \\
      & L3 & 40.00 & 45.00 & 47.50 & 52.50 & 43.36 \\
    \midrule
    \multirow{3}{*}{No iterative search}
      & L1 & 77.50 & 77.50 & 77.50 & 77.50 & 77.50 \\
      & L2 & 47.50 & 50.00 & 50.00 & 50.00 & 48.75 \\
      & L3 & 35.00 & 50.00 & 50.00 & 55.00 & 42.80 \\
    \midrule
    \multirow{3}{*}{No regional crop}
      & L1 & 87.50 & 90.00 & 90.00 & 90.00 & 88.75 \\
      & L2 & 62.50 & 62.50 & 65.00 & 67.50 & 63.28 \\
      & L3 & 42.50 & 55.00 & 60.00 & 65.00 & 50.00 \\
    \midrule
    \multirow{3}{*}{No OCR}
      & L1 & 77.50 & 77.50 & 77.50 & 77.50 & 77.50 \\
      & L2 & 50.00 & 55.00 & 57.50 & 62.50 & 53.90 \\
      & L3 & 52.50 & 55.00 & 65.00 & 67.50 & 55.82 \\
    \midrule
    \multirow{3}{*}{No page inspection}
      & L1 & 67.50 & 72.50 & 75.00 & 77.50 & 70.86 \\
      & L2 & 40.00 & 42.50 & 47.50 & 52.50 & 42.55 \\
      & L3 & 30.00 & 37.50 & 50.00 & 60.00 & 36.68 \\
    \midrule
    \multicolumn{7}{l}{\textbf{OCR-Text}} \\
    \midrule
    \multirow{3}{*}{Full agent}
      & L1 & 42.50 & 42.50 & 42.50 & 47.50 & 43.06 \\
      & L2 & 40.00 & 40.00 & 40.00 & 40.00 & 40.00 \\
      & L3 & 27.50 & 32.50 & 40.00 & 45.00 & 32.12 \\
    \midrule
    \multirow{3}{*}{No iterative search}
      & L1 & 30.00 & 30.00 & 30.00 & 30.00 & 30.00 \\
      & L2 & 22.50 & 25.00 & 25.00 & 25.00 & 23.75 \\
      & L3 & 30.00 & 32.50 & 37.50 & 37.50 & 32.08 \\
    \midrule
    \multirow{3}{*}{No regional crop}
      & L1 & 42.50 & 42.50 & 42.50 & 45.00 & 42.75 \\
      & L2 & 37.50 & 40.00 & 40.00 & 40.00 & 38.75 \\
      & L3 & 22.50 & 30.00 & 32.50 & 35.00 & 26.40 \\
    \midrule
    \multirow{3}{*}{No page OCR}
      & L1 & 40.00 & 40.00 & 40.00 & 42.50 & 40.36 \\
      & L2 & 30.00 & 30.00 & 30.00 & 32.50 & 30.31 \\
      & L3 & 30.00 & 35.00 & 37.50 & 42.50 & 33.22 \\
    \midrule
    \multirow{3}{*}{No page inspection}
      & L1 & 20.00 & 22.50 & 27.50 & 30.00 & 22.78 \\
      & L2 & 12.50 & 12.50 & 20.00 & 20.00 & 14.13 \\
      & L3 & 12.50 & 22.50 & 22.50 & 32.50 & 17.95 \\
    \bottomrule
  \end{tabular}
\end{table}

Table~\ref{tab:level-ablation} provides the level-wise ablations underlying the
aggregate results in Section~\ref{sec:ablations}.  Iterative search and page
inspection generally improve both routes, with inspection producing the largest
losses when removed.  Text access benefits L1 and L2 but has mixed effects on L3,
reflecting a path-dependent trade-off between local verification and corpus
exploration.  Cropping likewise has no stable effect across routes or levels.

\subsection{Results by Path Scope}
\label{app:path-scope-results}

\begin{table}[htbp]
  \centering
  \small
  \setlength{\tabcolsep}{5pt}
  \caption{GPT-5.6-sol results by evidence-path scope (\%).}
  \label{tab:scope-results}
  \begin{tabular}{llrrrrrr}
    \toprule
    \textbf{Route} & \textbf{Scope} & \textbf{Queries} & \textbf{R@1} & \textbf{R@3} & \textbf{R@5} & \textbf{R@10} & \textbf{MRR@10} \\
    \midrule
    \multirow{3}{*}{Visual}
      & Single-unit & 40 & 85.00 & 87.50 & 87.50 & 90.00 & 86.15 \\
      & Same-document & 74 & 52.70 & 56.76 & 58.11 & 60.81 & 55.19 \\
      & Cross-document & 6 & 16.67 & 16.67 & 16.67 & 16.67 & 16.67 \\
    \midrule
    \multirow{3}{*}{OCR-Text}
      & Single-unit & 40 & 42.50 & 42.50 & 42.50 & 47.50 & 43.06 \\
      & Same-document & 74 & 36.49 & 39.19 & 43.24 & 44.59 & 38.85 \\
      & Cross-document & 6 & 0.00 & 0.00 & 0.00 & 16.67 & 1.67 \\
    \bottomrule
  \end{tabular}
\end{table}

Table~\ref{tab:scope-results} separates single-unit, same-document, and
cross-document paths.  Visual R@1 decreases from 85.00\% on single-unit paths
to 52.70\% on same-document paths and 16.67\% on cross-document paths;
OCR-text R@1 decreases from 42.50\% to 36.49\% and 0.00\%, respectively.  The
scope breakdown shows that aligning evidence across documents remains the most
challenging setting, while visual retrieval retains an advantage in placing the
target first.

\section{Trace and Qualitative Analysis}
\label{app:trace-analysis}

\subsection{Target Progression Events}
\label{app:target-progression}

\begin{figure}[htbp]
  \centering
  \includegraphics[width=\linewidth]{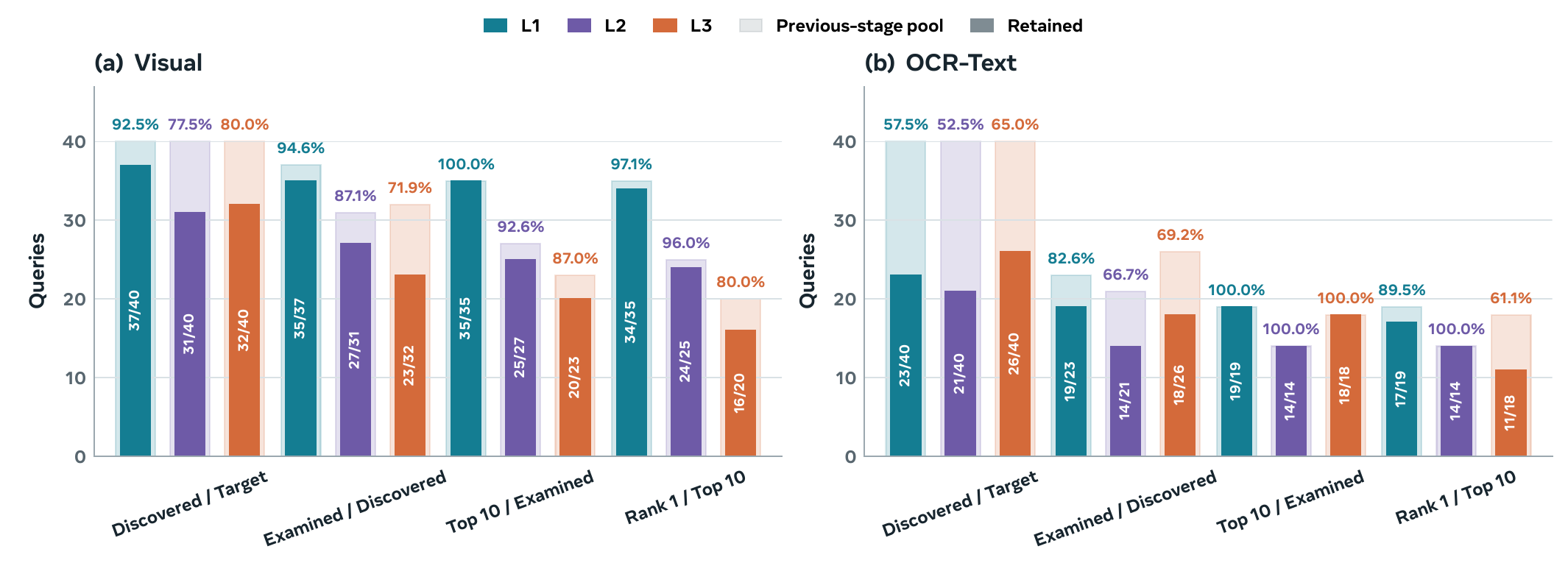}
  \caption{Level-wise gold-target progression through GPT-5.6-sol traces for
  (a) Visual and (b) OCR-Text retrieval. Colors distinguish L1, L2, and L3;
  pale bars show the previous-stage pool, while solid bars show the retained
  subset. Overall route rates appear in Figure~\ref{fig:target-progression}.}
  \label{fig:target-progression-levels}
\end{figure}

Our post-hoc audit uses planner responses, parsed actions, tool observations
including returned pages and crops, cumulative usage, and final rankings. We
map opaque handles back to annotation units after an episode finishes. The
overall progression in Figure~\ref{fig:target-progression} and its level-wise
breakdown in Figure~\ref{fig:target-progression-levels} use the following event
definitions:

\begin{itemize}
  \item The target is \emph{discovered} when its handle first appears in a
    successful search result at any rank.
  \item A full page is \emph{examined} at the first successful
    \texttt{inspect\_pages}, full-page \texttt{ocr\_image}, or
    \texttt{get\_page\_ocr} action involving it.  A successful crop also counts
    as examination of its source page.
  \item Target progression follows discovery, examination conditional on
    discovery, final top-10 inclusion conditional on examination, and rank-one
    placement conditional on top-10 inclusion.
\end{itemize}

\subsection{Two Audited Level-3 Trajectories}
\label{app:qualitative-trajectories}

\begin{figure}[htbp]
  \centering
  \includegraphics[width=\linewidth]{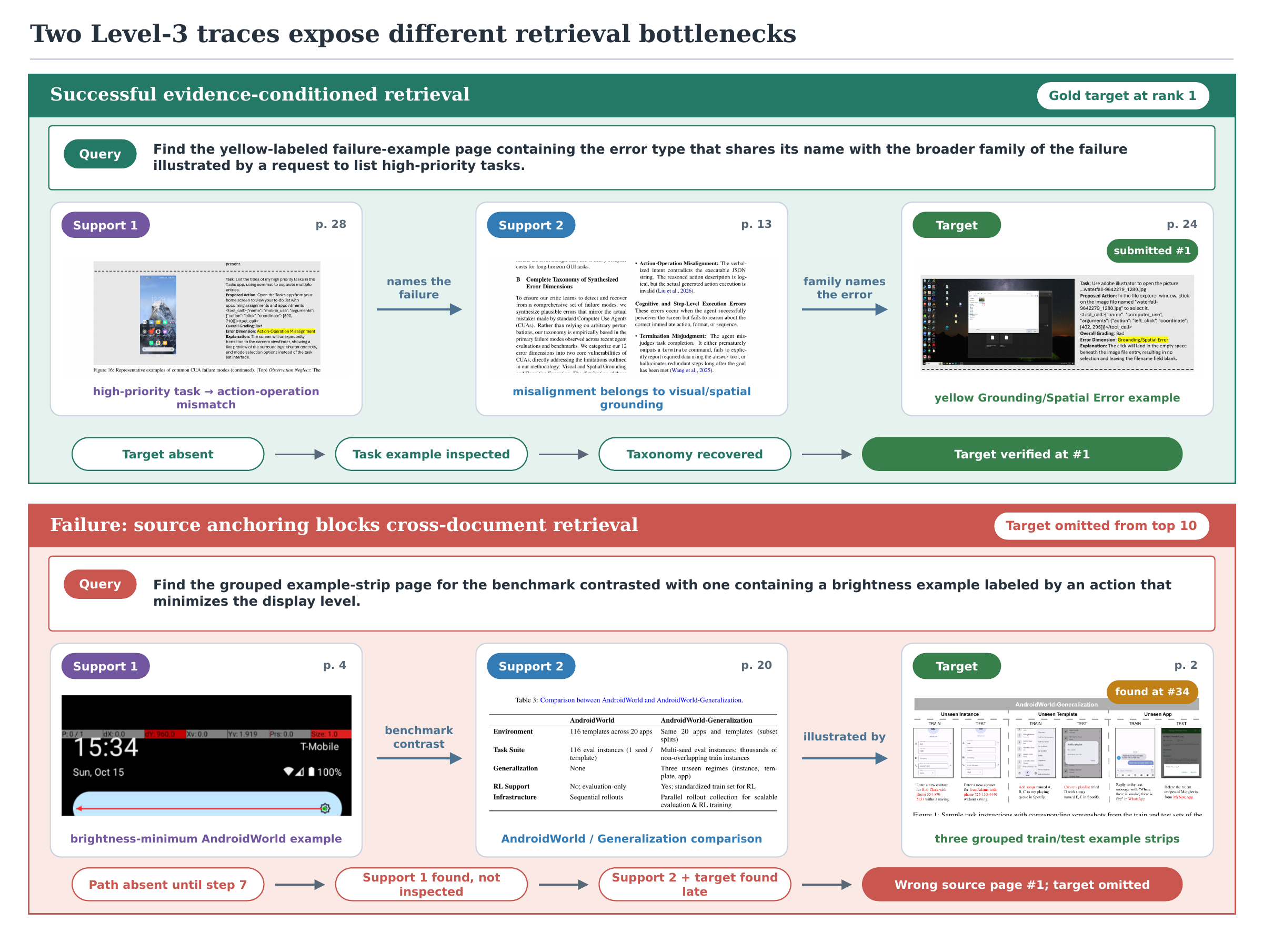}
  \caption{Queries and enlarged evidence regions for two audited Level-3
  visual-route episodes.  Top: the first search finds only the task-example
  support; inspecting it redirects retrieval toward the taxonomy and the
  yellow-labeled target, which is submitted at rank 1.  Bottom: the agent
  repeatedly anchors retrieval to the paper containing the brightness example.
  The benchmark-comparison support and target appear only in its final search
  at ranks 25 and 34, remain unexamined, and the target is omitted from the
  submitted top 10.}
  \label{fig:appendix-traces}
\end{figure}

\begin{table}[htbp]
  \centering
  \footnotesize
  \renewcommand{\arraystretch}{1.12}
  \setlength{\tabcolsep}{3.2pt}
  \caption{Complete successful visual-route trajectory for the top case in
  Figure~\ref{fig:appendix-traces}.  Search observations are summarized by the
  ranks of the annotated path pages.}
  \label{tab:trace-success}
  \begin{tabular}{c p{0.20\textwidth} p{0.25\textwidth} p{0.43\textwidth}}
    \toprule
    \textbf{Step} & \textbf{Decision rationale} & \textbf{Action} & \textbf{Observation and state change} \\
    \midrule
    1 & Search for the yellow-labeled examples and the high-priority-task failure. & \texttt{visual\_search}; top 30. & Support~1 appears at rank 2; Support~2 and the target are absent. \\
    \midrule
    2 & Inspect the strongest failure-example candidates. & \texttt{inspect\_pages} on ten pages, including Support~1. & The high-priority-task request is linked to Action--Operation Misalignment. \\
    \midrule
    3 & Search for that failure mode and its broader family. & \texttt{visual\_search}; top 30. & Support~1, Support~2, and the target appear at ranks 3, 6, and 7. \\
    \midrule
    4 & Refine toward a yellow-highlighted family label. & \texttt{visual\_search}. & The three path pages remain at ranks 3, 6, and 7. \\
    \midrule
    5 & Compare unseen pages in the failure-example series. & \texttt{inspect\_pages} on ten pages, including Support~2 and the target. & The taxonomy page and the yellow-labeled candidate are visually identified. \\
    \midrule
    6 & Return to the exact high-priority-task wording. & \texttt{visual\_search}; top 30. & Support~1 is rank 1, Support~2 rank 7, and the target rank 18. \\
    \midrule
    7 & Recheck the strongest task-example candidates. & \texttt{inspect\_pages} on ten pages, including Support~1. & No competing example supplies a better semantic chain. \\
    \midrule
    8 & Test a yellow-labeled alternative. & \texttt{ocr\_image} on a candidate page. & OCR identifies Action Formulation Error and Constraint Neglect, excluding it. \\
    \midrule
    9 & Test another visually matching alternative. & \texttt{ocr\_image} on a candidate page. & OCR identifies Suboptimal Path and Timing and Latency Neglect, excluding it. \\
    \midrule
    10 & Resolve the broader failure family. & \texttt{ocr\_image} on Support~2. & OCR places Action--Operation Misalignment under Visual and Spatial Grounding and names Grounding/Spatial Error. \\
    \midrule
    11 & Verify the namesake error on the target candidate. & \texttt{ocr\_image} on the target. & OCR confirms the yellow-highlighted Grounding/Spatial Error example. \\
    \midrule
    12 & Submit the verified target. & \texttt{submit\_answer} with a ranked top 10. & The target is rank 1, followed by Support~1 and Support~2. \\
    \bottomrule
  \end{tabular}
\end{table}

\begin{table}[htbp]
  \centering
  \footnotesize
  \renewcommand{\arraystretch}{1.12}
  \setlength{\tabcolsep}{3.2pt}
  \caption{Complete failed visual-route trajectory for the bottom case in
  Figure~\ref{fig:appendix-traces}.  Search observations are summarized by the
  ranks of the annotated path pages and the submitted rank-one page.}
  \label{tab:trace-failure}
  \begin{tabular}{c p{0.20\textwidth} p{0.25\textwidth} p{0.43\textwidth}}
    \toprule
    \textbf{Step} & \textbf{Decision rationale} & \textbf{Action} & \textbf{Observation and state change} \\
    \midrule
    1 & Search for grouped examples and a brightness-minimizing action. & \texttt{visual\_search}; top 20. & Neither support nor the target appears. \\
    \midrule
    2 & Inspect the strongest grouped-example candidates. & \texttt{inspect\_pages} on ten pages. & No annotated path page is examined, and no benchmark relation is recovered. \\
    \midrule
    3 & Combine benchmark-example and minimum-brightness cues. & \texttt{visual\_search}. & All three path pages remain absent. \\
    \midrule
    4 & Search the brightness instruction more literally. & \texttt{visual\_search}. & All three path pages remain absent. \\
    \midrule
    5 & Emphasize the grouped smartphone layout. & \texttt{visual\_search}. & All three path pages remain absent. \\
    \midrule
    6 & Inspect additional visual matches. & \texttt{inspect\_pages} on nine pages. & The inspected pages still provide no annotated path evidence. \\
    \midrule
    7 & Hypothesize AndroidControl or GUIOdyssey as the target benchmark. & \texttt{visual\_search}; top 30. & Support~1 first appears at rank 18; Support~2 and the target are absent. \\
    \midrule
    8 & Inspect candidates from the benchmark hypothesis. & \texttt{inspect\_pages} on ten pages. & Support~1 is not selected for inspection, so its AndroidWorld label remains unused. \\
    \midrule
    9 & Recast the query as a DragOn figure search. & \texttt{visual\_search}; top 50. & Support~1 rises to rank 4, but remains unexamined; Support~2 and the target are absent. \\
    \midrule
    10 & Search within DragOn using AndroidWorld and OSWorld cues. & \texttt{visual\_search}; top 50. & The eventual rank-one page is rank 2, Support~1 rank 4, Support~2 rank 25, and the target rank 34. \\
    \midrule
    11 & Prioritize the source-paper candidates. & \texttt{submit\_answer} with a ranked top 10. & A DragOn prose page is rank 1 and Support~1 rank 2; Support~2 and the target are omitted. \\
    \bottomrule
  \end{tabular}
\end{table}

We examine one successful and one failed GPT-5.6-sol visual-route episode.
Figure~\ref{fig:appendix-traces} enlarges the decisive regions of each support
and target page.  Tables~\ref{tab:trace-success}
and~\ref{tab:trace-failure} retain every planner decision and tool call, with
long search lists summarized by path-page ranks.

\paragraph{Evidence changes the successful search state.}
The first search in Table~\ref{tab:trace-success} finds the task example while
missing the taxonomy and target.  Inspection identifies its failure mode, and
the next search surfaces both remaining path pages.  Subsequent OCR resolves
the broader family and verifies its namesake error on the target page.  The
episode illustrates a complete transition from partial evidence to a grounded
rank-one decision.

\paragraph{Source anchoring can block a cross-document relation.}
The failed episode in Table~\ref{tab:trace-failure} retrieves no annotated path
page during its first six actions.  Once the search begins surfacing DragOn
pages, the planner increasingly treats that source document as the destination.
Support~1 reaches rank 4, and the final search exposes Support~2 and the target
at ranks 25 and 34, but none is examined.  The submission consequently ranks a
DragOn prose page first and omits the requested grouped example strip.  The
planner has mistaken the benchmark containing the brightness example for the
target instead of following the contrast relation to
AndroidWorld-Generalization, localizing the failure to relation-following and
candidate selection before visual verification.

\section{Licensing and Privacy}
\label{app:artifact-licensing}

We check the license attached to the exact arXiv version of every source
document. The corpus specification records that version, its source URL, and
the applicable license. Page images and substantial source-derived text are
redistributed only when the corresponding terms permit it. For other sources,
local corpus reconstruction proceeds from the versioned record using the
deterministic rendering procedure.

Benchmark-authored annotations and software are licensed independently of the
source documents. Their licenses do not extend to source text or images,
including excerpts present in OCR records or agent traces, which remain
governed by the corresponding source license.

Rendered scholarly pages may contain author names, affiliations, and contact
addresses already published in the source documents. We retain only this
source-visible content and the bibliographic metadata needed for source
identification and attribution; we neither infer personal attributes nor
combine the records with external personal data. Requests to correct or remove
source-derived personal information can be directed to the project contact.

\section{Open-Weight Baseline Resources and Runtime}
\label{app:open-weight-resources}

The open-weight planner is Qwen3.5-397B-A17B, with 397B total parameters and
17B activated parameters per token.  It is served in BF16 on two nodes, each
with eight NVIDIA A100-SXM4-80GB GPUs.  Tensor parallelism is eight within each
node and pipeline parallelism is two across nodes; evaluation uses four query
workers and the same 12-step protocol as the other planners.  The visual and
OCR-text routes use Qwen3-VL-Embedding-8B and Qwen3-Embedding-8B,
respectively.  PaddleOCR-VL-1.6 (0.9B) and PP-DocLayoutV3 provide page-text
recognition and layout detection.  Embedding and OCR services run on reserved
memory of the same local A100 infrastructure; construction of each static
embedding index uses one A100-SXM4-80GB GPU. All listed models use their
official weights.

Table~\ref{tab:qwen-runtime} summarizes the full 120-query inference time for
each combination of retrieval route and Qwen3.5 reasoning mode.

\begin{table}[htbp]
  \centering
  \small
  \setlength{\tabcolsep}{5pt}
  \caption{Approximate end-to-end inference time for each open-weight agent
  setting.  Timing starts after all model services are ready and excludes model
  loading and deployment.}
  \label{tab:qwen-runtime}
  \begin{tabular}{lc}
    \toprule
    \textbf{Qwen3.5 setting} & \textbf{Runtime for 120 queries} \\
    \midrule
    Visual, thinking & 4--5 hours \\
    Visual, no thinking & 1--1.5 hours \\
    OCR-Text, thinking & 6.5--7.5 hours \\
    OCR-Text, no thinking & 40--50 minutes \\
    \bottomrule
  \end{tabular}
\end{table}

\end{document}